\documentclass[a4paper,11pt]{article}
\usepackage{jcappub} % for details on the use of the package, please
\usepackage[T1]{fontenc} % if needed

\usepackage{tocloft}
\usepackage{natbib}
\usepackage{bm}
\usepackage{epsfig}
\usepackage{graphicx}
\usepackage{tikz}
\usepackage{pict2e}
\usepackage{amsmath}
\usepackage{mathptmx}
\usepackage{natbib}
\usepackage[scaled=.90]{helvet}
\usepackage{color}
\usepackage{gensymb}

\usepackage{natbib,twoopt}
\def\beq{\begin{equation}}
\def\eeq{\end{equation}}
\def\bfig{\begin{figure}}
\def\efig{\end{figure}}
\def\beqa{\begin{eqnarray}}
\def\eeqa{\end{eqnarray}}

\def\q{\mathbf q} 
\def\k{\boldsymbol k}
\def\x{\mathbf x}

\def\s{\mathbf s}
\def\v{\mathbf v}

\def\dbar{\overline{\delta}}

\def\sone{\s^{\scalebox{0.6}{(1)}}}
\def\sones{s^{\scalebox{0.6}{(1)}}}
\def\done{\delta^{\scalebox{0.6}{(1)}}}
\def\Dtwo{D^{\scalebox{0.6}{(2)}}}
\def\Dtwodot{\dot{D}^{\scalebox{0.6}{(2)}}}

\def\sonebar{\overline{\s}^{\scalebox{0.6}{(1)}}}
\def\stwobar{\overline{\s}^{\scalebox{0.6}{(2)}}}

\def\eone{{e}^{\scalebox{0.6}{(1)}}}

\def\dtwo{{\delta}^{\scalebox{0.6}{(2)}}}
\def\stwo{{\s}^{\scalebox{0.6}{(2)}}}
\def\nhat{\hat{n}}
\def\hhat{\hat{h}}

\def\sigT{\sigma_{\rm{T}}}

\makeatletter
\newcommand{\gsc}[1]{%
  \ifnum\c@section=0 %
  {{\bf\color{black} FFigReplace} -- {{\color{red}\textbf{CapBeg #1}}} Section="\thechapter"}
  \else
  \ifnum\c@subsection=0 %
  {{\bf\color{black} FFigReplace} -- {{\color{red}\textbf{CapBeg #1}}} Section="\thesection"}
  \else
  \ifnum\c@subsubsection=0 %
  {{\bf\color{black} FFigReplace} -- {{\color{red}\textbf{CapBeg #1}}} Section="\thesubsection"}
  \else
  {{\bf\color{black} FFigReplace} -- {{\color{red}\textbf{CapBeg #1}}} Section="\thesubsubsection"}
  \fi
  \fi
  \fi
}

\usepackage{enumitem}
\usepackage[percent]{overpic}

\usepackage{multirow}
\usepackage{booktabs}

\title{The Websky Extragalactic CMB Simulations}
\author[a,b,c,1]{George Stein,\note{Corresponding author.}}
\author[a]{Marcelo A. Alvarez,}
\author[c]{J. Richard Bond,}
\author[d,c]{Alexander van Engelen}
\author[e]{and Nicholas Battaglia}

\affiliation[a]{Berkeley Center for Cosmological Physics,\\341 Campbell Hall, University of California, Berkeley, CA 94720, USA}
\affiliation[b]{Lawrence Berkeley National Laboratory,\\Berkeley, CA 94720, USA}
\affiliation[c]{Canadian Institute for Theoretical Astrophysics, University of Toronto,\\60 St. George ST., Toronto ON, M5S 3H8, Canada}
\affiliation[d]{School of Earth and Space Exploration, Arizona State University,\\Tempe, AZ 85287, USA}
\affiliation[e]{Department of Astronomy, Cornell University,\\Ithaca, NY 14853, USA}

\emailAdd{gstein@berkeley.edu}

\abstract{We present a new pipeline for the efficient generation of synthetic observations of the extragalactic microwave sky, tailored to large ground-based CMB experiments such as the Simons Observatory, Advanced ACTPol, SPT-3G, and CMB-S4. Such simulated observations are a key technical challenge in cosmology because of the dynamic range and accuracy required. The first part of the pipeline generates a random cosmological realization in the form of a dark matter halo catalog and matter displacement field, as seen from a given position. The halo catalog and displacement field are modeled with ellipsoidal collapse dynamics and Lagrangian perturbation theory, respectively. In the second part, the cosmological realization is converted into a set of intensity maps over the range 10 - $10^3$~GHz using models based on existing observations and hydrodynamical simulations. These maps include infrared emission from dusty star forming galaxies (CIB), Comptonization of CMB photons by hot gas in groups and clusters through the thermal Sunyaev-Zel'dovich effect (tSZ), Doppler boosting by Thomson scattering of the CMB by bulk flows through the kinetic Sunyaev-Zel'dovich effect (kSZ), and weak gravitational lensing of primary CMB anisotropies by the large-scale distribution of matter in the universe. After describing the pipeline  and its implementation, we present the {\em Websky} maps, created from a realization of the cosmic web on our past light cone in the redshift interval $0<z<4.6$ over the full-sky and a volume of $\sim 600\ ({\rm Gpc}/h)^3$ resolved with $\sim\!10^{12}$ resolution elements. The Websky maps and halo catalog are publicly available at \href{https://mocks.cita.utoronto.ca/websky}{mocks.cita.utoronto.ca/websky}.}

\begin{document}
\maketitle
\flushbottom

\newpage
% ===========================
% Introduction
\section{Introduction}
\label{sec:intro}

The large-scale structure (LSS) of the Universe leaves its imprint on the cosmic microwave background (CMB) through a range of physical processes. Gravitational lensing, electron scattering, and emission from dust, stars, and ionized gas all trace out LSS and involve complex astrophysical phenomena associated with galaxies and the cosmic web, making the CMB an excellent probe of fundamental physics and galaxy formation. The cosmic infrared background (CIB) contains information about star forming galaxies over cosmic time; the thermal Sunyaev-Zel'dovich (tSZ) effect probes the energetics of galaxy clusters, groups, and massive galaxies, and can be used to determine their abundance; the kinetic Sunyaev-Zel'dovich (kSZ) probes velocity flows on large scales and the structure of the intergalactic medium (IGM) on small scales; and CMB lensing maps the total matter distribution of the universe projected along every line of sight. Taken together, these effects present powerful new ways to study inflation, modifications to general relativity, the mass of neutrinos, and galaxy formation, and are required to be well-understood to accurately reconstruct the primordial CMB. 

Using the CMB to probe LSS and galaxy formation presents unique challenges not present with intrinsically three-dimensional measurements such as with galaxy redshift and line intensity mapping surveys, however. The SZ effects and CMB lensing have a spectral dependence that is independent of the redshift distribution of the intervening structure, making them essentially two-dimensional, weighted projections. Although the CIB does contain some redshift information through its frequency dependence, it is difficult to extract due to uncertainties in the dependence of the rest-frame spectral energy distribution on galaxy mass, history, and environment. Simultaneously modeling these correlated effects, in order to disentangle them and extract information about the underlying structure, presents one of the most challenging problems in CMB data analysis \cite{2008A&A...491..597L,2011A&A...536A..18P,2018arXiv180706208P,2019arXiv191105717M}.  CMB maps at different frequencies can be correlated with each other and other LSS probes to learn about galaxy formation and obtain new constraints on cosmological model parameters. Prominent recent examples have used ACT, SPT, and Planck observations to find massive nearby galaxy clusters (e.g. \cite{2009ApJ...701...32S,2013JCAP...07..008H,2013ApJ...763..127R,2014A&A...571A..29P,2015ApJS..216...27B,2016A&A...594A..27P,2018ApJS..235...20H,2019arXiv190709621H}) and map out the full-sky of unresolved groups and clusters (e.g. \cite{2016A&A...594A..22P}) through the tSZ effect, detect large scale flows of the IGM through cross-correlation with galaxy redshift surveys (e.g. \cite{2016PhRvL.117e1301H,2016MNRAS.461.3172S,2017JCAP...03..008D}), model dusty star-forming galaxies with the CIB (e.g. \cite{2014A&A...571A..30P}), and to measure the clustering of matter through CMB lensing (e.g. \cite{2003PhRvD..67h3002O,2012ApJ...756..142V,2014PhRvL.113b1301A,2016ApJ...833..228B,2017PhRvD..95l3529S,2018arXiv180706210P,2019arXiv190505777W}).

Upcoming ground based CMB surveys, such as the Simons Observatory\footnote{\url{http://simonsobservatory.org}} \cite{2019JCAP...02..056A} and CMB-S4\footnote{\url{http://cmb-s4.org}} \cite{2016arXiv161002743A} to follow, will map out nearly half of the microwave sky at a resolution approaching an arcminute with unprecedented sensitivity, requiring correlated extragalactic foreground simulations with comparable resolution and sky coverage. Large simulation volumes are required in order to accurately capture large-scale velocity flows and to probe the statistics of massive rare objects such as high redshift luminous quasars, intermediate redshift proto-clusters, and the most massive galaxy clusters in our past light cone. Simultaneously, small scale effects associated with halo clustering and nonlinear structure within halos themselves must be modeled, particularly because non-Gaussian statistics associated with non-linearities couple long and short modes in non-trivial ways \cite{2014ApJ...786...13V}. The pipelines to generate these simulations should be efficient and repeatable, in order to (1) model the statistics of the underlying signal robustly by generating multiple Monte Carlo realizations and (2) study the effect of varying unknown astrophysical parameters over their plausible range, as well as beyond standard model effects such as modifications to gravity. 

Previous simulations modeled the different components with varying levels of accuracy, depending on the application, such as extragalactic point sources \cite{1997ApJ...480L...1G,1999ASPC..181..367J,2005ApJ...621....1G}, the tSZ effect in small regions (from individual clusters subtending a few arcmin to fields of view of $\sim$ 100 sq deg.) \cite{1988MNRAS.233..637C,2002ApJ...579...16W,2005MNRAS.356.1477D,2005ApJ...623L..63M,2006ApJ...650..538N,2006MNRAS.370.1309S,2007MNRAS.377..253M,2007ApJ...671...27H,2007MNRAS.378..385P,2007MNRAS.378.1259R,2008A&A...483..389P,2009MNRAS.397.2189P,2010ApJ...725...91B}, the CIB \cite{2008A&A...481..885F,2018arXiv181208575E}, the kSZ \cite{2016ApJ...823...98F}, and CMB lensing \cite{2008MNRAS.388.1618C,2015JCAP...03..049C,2017ApJ...850...24T}. State-of-the-art massively parallel cosmological simulations with hydrodynamics and subgrid feedback prescriptions still require considerable computational resources to generate even a single realization, and are currently limited to a maximum size of $\sim 1$~Gpc$^3$ (although generally much smaller) \cite{2014MNRAS.441.1270L,2015MNRAS.450.1349K,2016MNRAS.463.1797D,2016MNRAS.463.3948D,2017MNRAS.465.2936M,2017MNRAS.469.3069G,2018MNRAS.475..676S,2019ApJ...877...85E}.  Full-sky simulations that map out a volume corresponding to our past light cone therefore require approximate large-scale structure realizations without explicit hydrodynamic simulation of individual objects. Rather, small scale observations and hydrodynamic simulations can be used to calibrate efficient large-scale models by a halo model \cite{2012ApJ...758...75B} or the dark matter-only distribution \cite{2018JCAP...11..009D}. Large-scale cosmological N-body simulations focused on halo catalogs and the overall cosmic web of dark matter can be done much more quickly than high-resolution galaxy formation simulations, and were used as the input for a model to simulate the extragalactic microwave sky in \cite{2007ApJ...664..149S} and \cite{2010ApJ...709..920S}, including synchrotron and dust in galaxies, and lensing, with the resulting maps being used extensively in modeling of data from ACT and other CMB experiments to the present day. This approach of applying hydrodynamical results to N-body simulations was further validated in \cite{2012ApJ...758...75B}, where it was demonstrated that explicitly replacing pressure profiles in hydrodynamic simulations with the spherically-averaged mean for halos of a given mass and redshift produces similar results. 
 
An efficient new pipeline we have developed that builds on this previous work, and the publicly available sky maps generated with it, are the focus of this paper. The new pipeline combines (1) a parallel on-the-fly lightcone algorithm for fast halo catalogs with (2) a halo-field model for the observables, to generate simulations of the extragalactic sky from microwave to far infrared bands much more efficiently than previously possible. Among the available accelerated methods for generating realizations of large scale structure \cite{2013JCAP...06..036T,2016MNRAS.463.2273F,2002MNRAS.331..587M,2013MNRAS.428.1036M,2014MNRAS.439L..21K,2015MNRAS.450.1856A}, we have chosen to use the mass-Peak Patch approach \cite{2019MNRAS.483.2236S, 1996ApJS..103....1B} for this work. In \S\ref{sec:method} we describe the first step in the pipeline, efficient generation of a realization of LSS on the past light cone. In \S\ref{sec:mapmaking}
 we describe the second step, going from the LSS realization to maps of the sky at different wavelengths, from microwave to the far infrared. The results of the simulations compared to recent observational data are given in \S\ref{sec:skymocks}, including angular auto and cross power spectra of the CIB, tSZ, kSZ, and lensing signals. We conclude with a summary and discussion of future directions in \S\ref{sec:discussion}. The following flat $\Lambda$CDM cosmological parameters, consistent with Planck 2018 \cite{2018arXiv180706209P}, are used throughout this paper: $(\Omega_m,\Omega_b,\sigma_8,n_s,h,\tau)=(0.31,0.049,0.81,0.965,0.68,0.055)$. 

% ===========================
% Peak Patch Method
\section{Large Scale Structure Simulations}

\label{sec:method}

The sky fraction, redshift coverage, and mass resolution required for accurate extragalactic mocks of current and next-generation CMB observations demand exceedingly large simulations. For example, a single-box full-sky lightcone simulation out to $z=5$, capable of resolving dark matter halos above a minimum mass of $\sim$1$\times$ 10$^{12}$ M$_{\odot}/h$ with 20 resolution elements each, requires at least $\sim$12,000$^3$ total resolution elements in a box size of $\sim$10 Gpc$/h$. This will remain computational infeasible for hydrodynamical methods over the next decade, and a single realization would be currently achievable through full N-body simulation only at the expense of many millions of CPU hours -- a considerable investment for all but the largest computational grants. One approach to alleviate this computational bottleneck is to instead simulate a smaller volume with either hydrodynamical or N-body methods and replicate it as needed to fill the sky fraction and redshift requirements. Unfortunately, this repeats structures along the line of sight, does not reproduce the statistics of rare objects, and fails to capture the effects of the large-scale density and velocity fluctuations that are important for accurate mocks of gravitational lensing and the kSZ effect.

The demand for rapid generation of these type of simulations has therefore resulted in the development of many `approximate' methods for generating realizations of large scale structure \cite{2013JCAP...06..036T,2016MNRAS.459.2327I,2016MNRAS.463.2273F,2002MNRAS.331..587M,2013MNRAS.433.2389M,2013MNRAS.428.1036M,2014MNRAS.439L..21K,2015MNRAS.450.1856A,2019MNRAS.483.2236S,2019MNRAS.482.2861B,2019PNAS..11613825H}. Many of these simulation methods were recently featured and discussed in a comparison project within the Euclid collaboration focused on determining their viability for estimating covariance matrices \citep{2019MNRAS.482.1786L, 2019MNRAS.485.2806B, 2019MNRAS.482.4883C}. This work provided an extensive analysis of the halo correlation function, halo power spectrum, and halo bispectrum, and their individual covariances, from a set of 300 approximate simulations at $z=1$, in comparison to the results of full N-body. To generate the dark matter halo catalogues for this work we use the mass-Peak Patch approach, recently described and validated in detail in \cite{2019MNRAS.483.2236S} and included in the Euclid comparison project. The mass-Peak Patch approach has been extensively compared to more expensive $N$-body simulation runs and alternative approximate methods, with highly-satisfactory results at the simulation resolutions required for this work. Individual peak-to-group comparisons show good agreement for high-mass, tightly bound groups, with growing scatter for lower masses and looser binding. The final state (Eulerian) spatial distribution of peak patches and N-body clusters, and halo velocities, have been shown to be satisfyingly close. As well, higher order halo statistics such as the halo bispectrum have been shown to be accurately reproduced when considering abundance-matched samples \cite{2019MNRAS.482.4883C} such as presented here. We use mass-Peak Patch in this work because it is computationally inexpensive, generates light cone catalogs on the fly with no explicit time-stepping, and the method fared well in the Euclid comparison project and subsequent validations. Using the same sky model applied in this work on a large scale structure realization generated using a different approximate simulation method should produce statistically similar extragalactic maps. 

The cosmological realizations presented here include halos and the material exterior to halos, the latter of which we refer to as the `field' component.  The halo properties are determined using the mass-Peak Patch approach \cite{2019MNRAS.483.2236S, 1996ApJS..103....1B}, while the dynamics of the field component, mainly of applicability for creating kSZ and lensing maps, is based on second order Lagrangian perturbation theory (2LPT) \cite{1995A&A...296..575B}. We provide a brief summary of the mass-Peak Patch method below, but refer the reader to \cite{2019MNRAS.483.2236S} for detailed explanations and validation results. The light cone pipeline can be separated into four main subprocesses: 
\begin{enumerate}
    \item Generation of a linear random Gaussian density field, $\delta$, and the corresponding 2LPT displacement vectors $\sone$ and $\stwo$: \S\ref{sec:ICs}
    \item Calculate candidate peak collapse dynamics under the homogeneous ellipsoid approximation to find potential collapsed regions: \S\ref{sec:peak-patch}
    \item Exclusion and merging of the collapsed regions in Lagrangian space to determine the final halo catalogue: \S\ref{sec:exclusion}
    \item  Assignment of 2LPT displacements and velocities to the halo and matter distribution: \S\ref{sec:exclusion}
\end{enumerate}

% ---------  ---------  ---------  ---------  ---------  ---------  --------- 
\subsection{Initial Conditions \& Domain Decomposition}
\label{sec:ICs}
The initial conditions for a mass-Peak Patch simulation, as is standard for the vast majority of large-scale structure simulations, are in the form of a cubic periodic lattice of any specified physical size $L$ and one dimensional resolution $n$. The total number of lattice sites, or equivalently particles, in the simulation is then $N=n^3$, initially uniformly distributed in a total volume of $V=L^3$. The uniform periodic lattice allows many operations to be performed in the Fourier domain to accelerate computation of convolutions and derivatives, where here we define the relation of a Lagrangian field $f(\q)$ and its Fourier transform $f(\k)$  as $f(\q) = (2\pi)^{-3} \int d\k e^{i \k \cdot \q} f(\k)$.

To generate the initial conditions on a periodic lattice, random perturbations are realized by first generating a white noise field by drawing each lattice value from a Gaussian distribution with a unit variance and a mean of zero. It is then convolved with the linear matter power spectrum to obtain the linear density contrast $\delta(\k)$. The first and second order linear displacements, describing the  displacement of matter as a function of time and Lagrangian position, are obtained using 2LPT, requiring the storage of seven values at each lattice site: \{$\delta(\q),\ \sone(\q),\  \stwo(\q)\}$. All quantities are expressed in terms of their values at $z=0$, and scaled back in time by the appropriate linear growth factor where required, unless otherwise noted.

The parallelization of the subsequent peak finding algorithm and homogeneous ellipsoid calculations are based on a domain decomposition consisting of overlapping cubic `tiles', which greatly reduces the need for communication and collective operations in the peak finding and exclusion algorithms. This choice allows maximum flexibility in separating the cosmological fluctuations into long and short-range components, with the possibility for a coarser resolution for long wavelength modes and adaptive resolution of smaller structures. For the full-sky simulations presented in this paper, uniform spatial resolution over a spherical region is required, and so the initial conditions are realized at uniform resolution on the periodic cubic lattice. Convolutions that require collective communication, such as calculation of displacements from the linear density field using perturbation theory, are done globally in parallel at the native resolution of the realization with the use of the slab-decomposed FFTW3 library\footnote{http://www.fftw.org/}. Depending on memory constraints, the density and displacement fields are stored either in main memory with a slab-decomposition, or as segmented files on disk for later processing. 

Once initial condition generation is complete, exchange is performed between processes to go from a slab to cubic domain decomposition with $N_{\rm t}^3$ tiles. For example, an eight-process run would start by realizing the linear density contrast and displacements in eight slabs, one for each processor. After this initial step, the fields would be rearranged on the processes such that there are eight cubic tiles in a 2 x 2 x 2 configuration, $N_{\rm t}=2$. Because the subsequent peak finding in each tile is done in parallel without any communication between tiles, including smoothing the density and displacements with a spherically-symmetric filter, it is necessary to retain information about the field in the nearby adjacent tiles through the use of a buffer region. The buffer region is taken to have a width given by the largest possible halo radius in Lagrangian coordinates. For example, in a Gpc-size region at $z=0$, one can expect to find halos as large as several times $10^{15}\ {\rm M}_\odot$, corresponding to a Lagrangian radius of about 40 Mpc. The smoothing can then be done locally using a threaded FFT on the tile with no further domain decomposition. For further details of the parallelization scheme and computational requirements see Appendix A of \cite{2019MNRAS.483.2236S}.

% ---------  ---------  ---------  ---------  ---------  ---------  --------- 
\subsection{Ellipsoidal Collapse}
\label{sec:peak-patch}

The halo catalogs are based on the parallel implementation of the mass-Peak Patch approach \cite{2019MNRAS.483.2236S}, wherein ellipsoidal collapse is used to determine locations and masses of potential collapsed objects (halos). The mass-Peak Patch approach is a Lagrangian space halo finder that associates halos with the largest regions that have just collapsed by a given redshift or, equivalently, distance from the observer. The determination of whether any given region will have collapsed or not is made by approximating it as a homogeneous ellipsoid, the fate of which is determined completely by the principal axes of the deformation tensor of the linear displacement field  (i.e. the strain) averaged over the region. In principle, the process of finding these local mass peaks would involve measuring the strain at every point in space, smoothed on every scale. However, experimentation has shown that equivalent results can be obtained by measuring the strain around {\em density} peaks found on a range of scales. This is not to say that a halo found on a given scale corresponds to a peak in the density smoothed on that scale, however, which is only the case when the strain is isotropic and the collapse is spherical. Thus, the use of density peaks as centers for strain measurements and ellipsoidal collapse calculations in the algorithm is only an optimization, to avoid wasting computations measuring the properties of regions of Lagrangian space that will not collapse in the first place.

The first step in identifying the possible site of a halo is therefore to identify peaks in the linear density field. This is done by smoothing the field on a series of top hat filters of different scales, which we will refer to as the `filter bank'. The filter bank can consist of logarithmically-spaced filter radii $R$, or linear of logarithmic spacing in $\sigma(R)$, with optimal filter spacings to maximize both accuracy and efficiency presented in \cite{2019MNRAS.483.2236S}, from a minimum radius of $R_{\rm f,min} = 2 a_{\rm latt}$, where $a_{\rm latt}$ is the lattice spacing, to a maximum radius of the largest halo expected to be found in the simulated volume ($R_{\rm f,max}=36 $~Mpc at $z=0$). Starting with the largest smoothing scale, a local FFT is performed on the local tile in order to convolve the field with the top hat smoothing kernel ${W}(kR) = 3j_1(kR)/(kR)$ where $j_1$ is the spherical Bessel function of order one. All points that are density peaks on that filter scale with a density contrast above a threshold determined from spherical collapse, roughly 1.5, are added to the list of candidate halo centers with a `peak radius' given by the scale at the filter scale, $R_{\rm pk}=R_{\rm f}$. This process is repeated for each smoothing scale in the filter bank until a full list of candidate halo centers has been compiled.

Once the candidate list is completed, measurements are made by averaging the linear strain, $e_{ij}\equiv -(\partial{\sones_i}/\partial{q_j}+\partial{\sones_j}/\partial{q_i})/2$, over spheres with Lagrangian radii $R$, centered on each halo candidate. The measurements are started from some initial radius chosen to be a fixed multiple of the smoothing scale on which the candidate center was found to be a density peak $R_{\rm init}=f_{\rm init}R_{\rm pk}$. If the region averaged over the initial radius is determined to collapse, the measurements are performed at progressively larger radii until the region no longer collapses, while if the region averaged over the initial radius does not collapse, the measurements are performed at decreasing radii until the region collapses, or is equal to a radius smaller than a lattice size. The mean strain within a given measurement radius is diagonalized to obtain its eigenvalues, $\lambda_i=-(\done\!/3)(1+c_i)$, and eigenvectors, $\nhat_i$, with $e_{ij}=\sum_k \lambda_k\nhat_k^i\nhat_k^j$, 
 $c_1=p+3e$, $c_2=-2p$, and  $c_3=p-3e$, where the linear density contrast and strain, $\done$ and $e_{ij}$, respectively, are averaged over the smoothing radius $R$. 
The redshift at which the corresponding homogeneous ellipsoid would collapse is calculated from the following system of ordinary differential equations:
\beq
\frac{\ddot{x}_i}{x_i}= \frac{\ddot{a}}{a} - \frac{1}{2}\Omega_mH^2\left[b_i\delta+c_i\delta_{\rm lin}\right],
\label{eq:ellipsoid}
\eeq
%\beq
%\frac{\ddot{x}_i}{x_i}= \frac{\ddot{a}}{a} - \frac{1}{2}\Omega_mH^2\left[b_i\dbar+\delta_{\rm lin}-3\lambda_i\right],
%\label{eq:ellipsoid}
%\eeq
where $x_i(t)=R_i(t)/R_{\rm m}$ are the scale factors in each of the three principal axes of the ellipsoid, $a$ is the background scale factor, $\delta=a^3/(x_1x_2x_3)-1$ is the evolving non-linear density contrast of the homogeneous ellipsoid, $\delta_{\rm lin}=\done{D}(a)$, and 
\beq
b_i(t)\equiv  \frac{3}{2}\int_0^\infty d\tau \left[\tilde{x}_i^2 +\tau\right]^{-1}\prod_j\left[\tilde{x}^2_j+\tau\right]^{-1/2},
\eeq 
where $\tilde{x}_i\equiv x_i/(x_1x_2x_3)^{1/3}$. The parameters $b_i$ are defined by the Newtonian potential of an isolated homogeneous ellipsoid in a coordinate system aligned with its principal axes: $\Phi(\x)=4\pi{G}\rho\sum_i b_ix_i^2/2$, and depend only on its evolving shape. Therefore, the first term in the brackets on the right hand side of equation (\ref{eq:ellipsoid}) is proportional to the gravitational acceleration from material inside of the ellipsoid, and is exact, while the second term represents that from external tidal fields, and is approximated by the linear solution, which only depends on the mean linear strain within $R$, through the $c_i$ \cite{1996ApJS..103....1B}. The parameters $b_i$ and $c_i$ can be considered to encode the anisotropy of the internal and external forces, respectively. Since the shape of the ellipsoid is time dependent, while the shape of the linear tidal field is not, $b_i$ is time dependent while $c_i$ is not. Note that in the spherical case, $x_1=x_2=x_3=x$ implies $b_i=1$, $c_i=0$, and $\ddot{x}/x = \ddot{a}/a - \Omega_mH^2\dbar/2$, which is the usual equation of motion for spherical collapse. 
The initial conditions at time $t_0$ for these equations are obtained using perturbation theory. Each axis of the ellipsiod is evolved until it reaches a critical radius $x_{eq,i} = f_{r,i} a$ during its collapse, after which it is `frozen in' at that value. Since the eigenvalues of the ellipsoid satisfy $\lambda_{v3} \geq \lambda_{v2} \geq \lambda_{v1}$, the 1-axis will be last to collapse, and it is at this point when a peak is considered virialized. The radial freezeout factors $f_{r,3}$ and $f_{r,2}$ were chosen to be 0.171, as this corresponds to the standard top hat virial density contrast of 200. The final axis was frozen out at $f_{r,1}=0.01$, indicating complete collapse along the final axis.

For each candidate peak we find the largest radius for which a homogeneous ellipsoid with the measured mean strain in the sphere contained within the radius would collapse by the redshift of interest, $z=z_{\rm coll}(\delta_m,e_m,p_m)$, with the collapse redshift function $z_{\rm coll}$ stored in a precomputed table for computational efficiency. If a candidate peak has no radius for which a homogeneous ellipsoid with the measured strain would have collapsed, then that point is discarded. Each candidate position and radius is then stored as a \textit{peak patch}. We then proceed down through the filter bank and repeat this procedure for each scale, resulting in a list of peak patches which we refer to as the `un-merged catalog'.

% ---------  ---------  ---------  ---------  ---------  ---------  --------- 
\subsection{Exclusion and Displacements}
\label{sec:exclusion}
Exclusion is essential to avoid double counting of matter in halos, since distinct halos do not overlap, by definition. Because neighboring and overlapping regions in Lagrangian space undergo ellipsoidal collapse in our calculations, we must use some practical algorithm for ensuring that exclusion is taken into account. The approach here is essentially unchanged from ``binary exclusion'' as defined in section 3.2 of \cite{1996ApJS..103....1B}. 

Binary exclusion starts from a  list of candidate peak patches sorted by mass or, equivalently, Lagrangian peak patch radius. For each patch we consider every other less massive patch that overlaps it. If the center of the smaller patch is inside the large one, then that patch is removed from the list. If the center of the smaller patch is outside of the larger one, then the volume of overlap of the two intersecting spheres is calculated from the standard equations describing the volume of their overlapping spherical caps. Considering the plane perpendicular to the separation vector between them, we subtract any volume beyond this plane from each of the halos. The overlapping mass from all neighbouring halos is accounted for before reducing their volumes. This process is repeated until the least massive remaining patch in the list is reached.

We use Lagrangian perturbation theory (LPT) to move halos and the matter outside them from their initial Lagrangian positions to their final Eulerian positions on the light cone. The first step is to compute the 1LPT and 2LPT displacements for each of the mass elements in the simulation, which is done globally in parallel when generating the initial conditions. The 1LPT displacements are obtained from $\sone(\k)=-i\done(\k)\k/k^2$ and the second order displacements are obtained with $\stwo(\k)=-i\dtwo(\k)\k/k^2$, where $\dtwo(\q)=\sum_{i>j}[
\eone_{ii}(\q)\eone_{jj}(\q)-
\eone_{ij}(\q)\eone_{ij}(\q)]$. In practice, we perform this sum by calculating the elements of the strain tensor in harmonic space, $\eone_{ij}(\k)=-k_ik_j\done(\k)/k^2$, and Fourier transforming to configuration space and performing the multiplications necessary sequentially, which maintains the memory usage of the code at seven floats per mass element ($\done$, $\sone$, $\stwo$), at the expense of only one additional Fourier transform. During the adaptive measurement step to determine the largest collapsing region around candidate peak patch centers, we also keep track of the mean displacements, $\sonebar$ and $\stwobar$, so that each unmerged object in the catalog is associated with its mean LPT displacement. 

% ---------  ---------  ---------  ---------  ---------  ---------  --------- 
\subsection{Light Cone}
\label{sec:lightcone}
The algorithm we have presented here can be applied at fixed time or from a fixed vantage point -- i.e., on the past light cone of the observer, which is the case for all the simulations we present here. Common choices for the observer are either at the center of the periodic simulation volume, with side $\lambda_{\rm box}$, or placed at a corner to consider only one octant of the sky. Every comoving cell in the simulation volume, at position $\q_{\rm c}$, corresponds to a fixed redshift given a cosmological model, $z_{\rm c}=z(\chi_{\rm c})$, where $\chi_{\rm c} = |\q_{\rm c}|$.   Each halo and field mass element is moved to its Eulerian position using $\x(z_c)={\q} + D(z_c)\sone + \Dtwo(z_c)\stwo$, where we use $\Dtwo(z) = 3/7D^2(z)\Omega_m^{-1/143}$, which is an excellent approximation \cite{1995A&A...296..575B}. 

%============================
% Map making
\section{Sky Model}
\label{sec:mapmaking}

The simulated extra-galactic microwave sky maps presented in this work are generated from the cosmological realizations by projecting, along the line of sight to each pixel and frequency, the emission from dusty star forming galaxies (CIB), Comptonization of CMB photons by hot gas in groups and clusters (tSZ), Doppler boosting by Thomson scattering of the CMB by bulk flows (kSZ), and weak lensing of the CMB by the intervening matter distribution. In \S\ref{subsec:general-skymodel} we first describe the general use of diffuse halo profiles, halo occupation distribution models, and treatment of field particles used, while \S\ref{subsec:astro-model} details the specific astrophysical models applied.

% ---------  ---------  ---------  ---------  ---------  ---------  --------- 
\subsection{General Approach}
\label{subsec:general-skymodel}

%%%%%%%%%%%%%%%%%%%%%%%%%%%%

The starting point for generating sky maps is the large scale structure along the observed lightcone, i.e. the mass and positions of halos and the LPT displacement field.  For lensing and SZ secondaries, we use continuous density and pressure profiles, while for the CIB we approximate group and cluster galaxies as point-sources with a stochastic halo occupation distribution (HOD) model. The pipeline has been developed in such a way that other models, such as oriented anisotropic profiles that depend on the tidal tensor and more complicated dependence on the assembly history of halos and their environments, are straightforward to implement. A halo-based approach is not sufficient to include the large scale correlations induced by lensing and the kSZ effect, which have a non-negligible contribution from regions exterior to galaxy and cluster scale dark matter halos. We hereafter refer to all material outside of halos resolved by the simulation as the `field' component.  We use second-order Lagrangian perturbation theory to predict the electron momentum and total matter density of this field component, ensuring that large-scale kSZ and lensing correlations are included self-consistently. 

\begin{figure}
\begin{center}
\includegraphics[width=0.9\textwidth, trim = 0 -20 0 0 0]{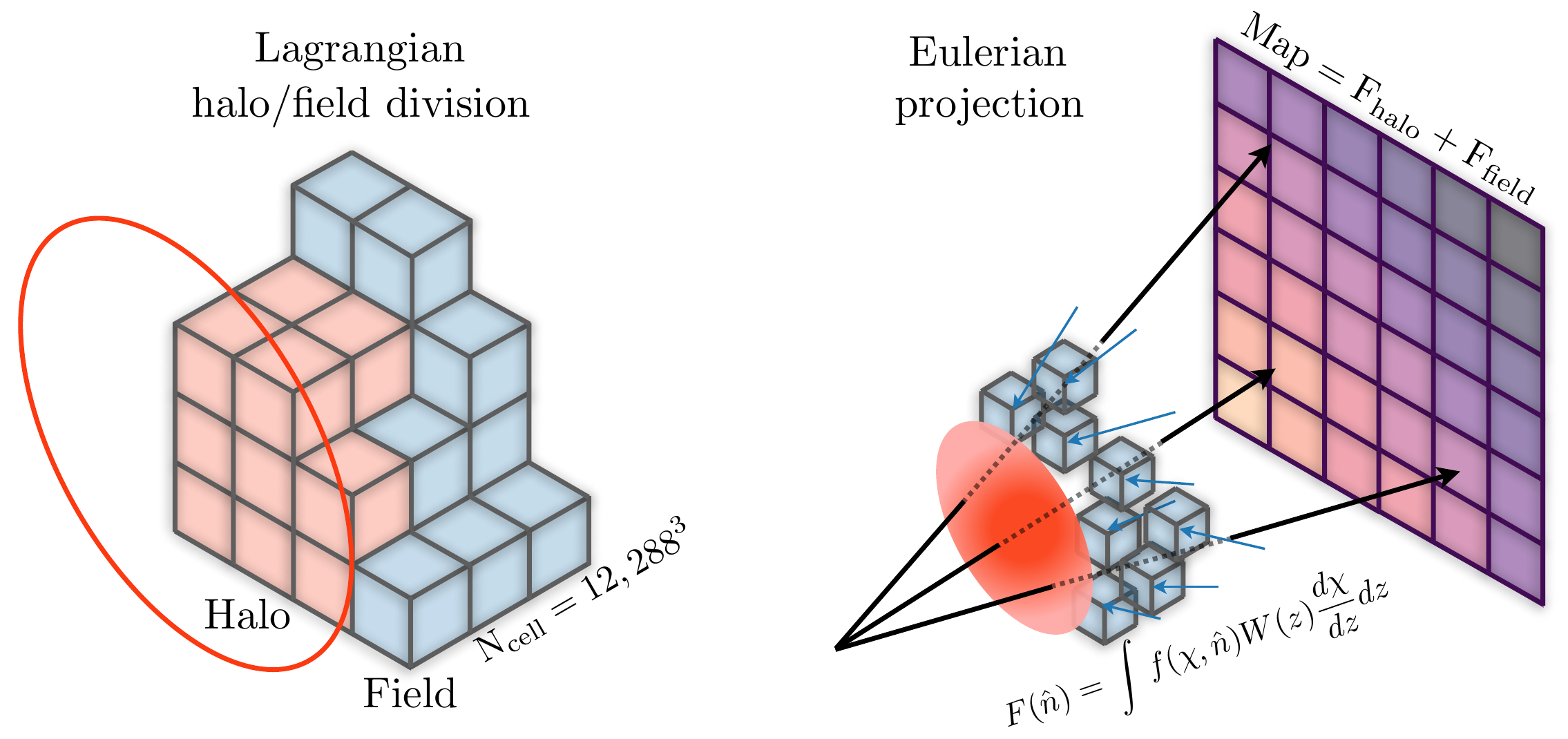}
\vspace{-0.8cm}
\caption{{\em left} -- Illustration of the separate Lagrangian treatment of simulation cells (particles) interior and exterior to halos.  Material interior to halos is shown in red, and the field is shown in blue, where the dynamics of both the field and halos are determined by LPT. Material inside halos is redistributed over either a diffuse radial profile (Section~\ref{subsubsec:gen_diffuse}), as shown in red, or into distributions of subhalos (Section~\ref{subsubsec:gen_hod}). The contribution from each field particle is determined by the field redshift-volume kernel function described in  Section~\ref{subsubsec:gen-exterior}.}
\end{center}
\label{fig:field-halo}
\end{figure}

We illustrate the separate treatment of simulation cells, or particles, interior and exterior to halos identified in this work with the mass-Peak Patch method in Figure~\ref{fig:field-halo}. Material interior to halos is shown in red, and the field is shown in blue. The dynamics of both field and halo are treated by calculating displacements using Lagrangian perturbation theory. These displacements are used to move the field particles and halos to Eulerian space. Material inside halos is redistributed over either a diffuse radial profile (Section~\ref{subsubsec:gen_diffuse}), as shown in red, or into distributions of subhalos (Section~\ref{subsubsec:gen_hod}). The contrubution from each field particle is determined by the field redshift-volume kernel function described in  Section~\ref{subsubsec:gen-exterior}.

Full-sky component maps for the halo component are generated using the `scatter' approach for the diffuse halo profiles, where each pixel is assumed to be smaller than the angular variation across the halo, while field and CIB maps are generated using a `gather' approach, where the contribution from a given mass element or galaxy is treated as a point and therefore contributes to only a single pixel. In the case where either of these limits is not satisfied, pixels or lattice sites can be subdivided, depending on requirements, which is particularly easy to do with regular sky pixelizations such as Healpix \cite{2005ApJ...622..759G,2019JOSS....4.1298Z} and the regular periodic lattices used to characterized the field displacements. 

Generically, weak lensing convergence $\kappa$, kSZ temperature fluctuation $\Delta{T}/T$, and tSZ Compton $y$-parameter $y$ in a given direction $\nhat$ can each be expressed as a specific quantity, $f$, integrated along the line of sight, 
\beq
F(\nhat)=\int f(\chi,\nhat)W(z)\frac{d\chi}{dz}dz,
\eeq
where $W(z)$ is redshift-dependent kernel that encapsulates the physics associated with $F$. Since all the effects considered here are linear, the final map is a sum over individual components, 
\beq
\label{eq:totalmap}
F(\nhat)=F_{\rm halos}(\nhat)+F_{\rm field}(\nhat)+F_{\rm point-sources}(\nhat),
\eeq
for diffuse halo profiles, field, and point-sources, respectively. A schematic diagram of this separate treatment of halos and the field is shown in Figure~\ref{fig:field-halo}.

%BEGIN FIGURE -------------

\subsubsection{Diffuse Halo Profiles}
\label{subsubsec:gen_diffuse}
 Once a halo light cone catalog has been created we use the linearity of equation \ref{eq:totalmap} to optimize the generation of halo maps through the use of look-up tables. In the cases we present here, these tables specify the contribution from a halo with a given mass $M$, redshift $z$, and angular position $\hat{h}$, to the map at a given point in the sky $\nhat$:
\beq
I_h(\mu|M_h,z_h) = R_h\int_{x_\mu}^{x_0} f_h(x|M,z)\left[1-x^2_\mu/x^2\right]^{-1/2}dx,
\eeq
where $\mu\equiv \nhat\cdot\hhat_h$, $x_\mu^2 \equiv \chi^2(z)(1-\mu^2)/(\mu^2R_h^2)$, and $f_h(r/R_h|M,z)$ encodes the complex astrophysics associated with, e.g., the matter or pressure distribution. All the diffuse halo profiles models adopted in this paper have the form of a generalized NFW profile (gNFW; \cite{1996MNRAS.278..488Z}) with a mass and redshift dependence parameterized by
\beq
f_h(x|M,z) =  f_0(M,z)
   \left[\frac{x}{x_c(M,z)}\right]^{\gamma(M,z)}\left\lbrace 1 +
   \left[\frac{x}{x_c(M,z)}\right]^{\alpha(M,z)} \right\rbrace^{-\beta(M,z)}.
   \label{eq:gnfw}
\eeq
For each halo $i$ at angular position ${\hhat}_i$ in the catalog, its contribution to each pixel $j$ is given by
\beq
\delta{F}^{\rm h}_{ij} \equiv \frac{1}{\delta\Omega_j}\int_{\Omega_j}I_h(\nhat\cdot\nhat_i|M_i,z_i)d\nhat,
\eeq
where the integration is over the solid angle subtended by the pixel, with $\delta{\Omega_j}\equiv \int_{\Omega_j}d\nhat_j$. We evaluate the values sampled at the pixel center to obtain an initial estimate of the value using 
\begin{equation}
\delta{F}^{\rm hs}_{ij} \equiv F_h(\nhat_i\cdot\nhat_j|M_i,z_i),
\end{equation}
and ensure the integrated contribution of the halo over all overlapping pixels is correct by an overall normalization factor:
\begin{equation}
\delta{F}^{\rm h}_{ij} = 2\pi\frac{\delta{F}^{\rm hs}_{ij}}{\sum_i\delta{F}^{\rm hs}_{ij}}\int_{-1}^1 d\mu I_h(\mu|M_i,z_i).
\end{equation}
The total value for each pixel $i$ in the halo map, $F^{\rm h}_{i}$, is obtained by summing over all halos, $F^{\rm h}_{i} = \sum_j \delta{F}^{\rm h}_{ij}$. The values of $j$ for which $\delta{F}_{ij}\neq 0$ for a given $i$ are computed using the condition 
\begin{equation}
(\hat{n}_i\cdot\hat{n}_j)^2 >\chi^2(z_j)/\left[\chi^2(z_j)+x_0^2R^2_h(M_j,z_j)\right].
\end{equation}

We convert between peak patch halo masses and spherical overdensity masses fit in the hydrodynamical simulations by abundance matching along the light cone using the universal halo mass function of \cite{2008ApJ...688..709T} with a mass 200 times the mean, $M_{\rm 200m}$. When necessary, we convert to halo masses with a different overdensity by assuming an NFW profile with a fixed concentration, $c = 7$, to be consistent with the mass profiles used in the lensing calculation (see \S\ref{subsubsec:cmb-lensing}).

%%%%%%%%%%%%%%%%%%%%%%%%%%%%
\subsubsection{Point Source Halo Occupation Distribution}
\label{subsubsec:gen_hod}

We assume galaxies populating a given halo emit with the same spectral shape, separated into central and satellite galaxies. The distribution of subhalos of a given mass $m$, hosted in halos with mass $M$, is used to assign luminosities to satellites by assuming that each satellite's properties are determined by the subhalo mass and redshift.  In particular we use the fit of subhalo mass function (SHMF) from N-body simulations in \citep{2014MNRAS.440..193J},
\begin{equation}
 \frac{dN_{\rm sh}}{d\ln{m}}(m|M) =  \left [   \gamma_1\left(\frac{m}{M}\right)^{\alpha_1} + \gamma_2\left(\frac{m}{M}\right)^{\alpha_2}\right ]\exp \left [  -\beta\left(\frac{m}{M}\right)^\zeta \right ],
 \label{eq:shmf}
\end{equation}
with $(\gamma_1, \alpha_1, \gamma_2, \alpha_2, \beta, \zeta) = (0.13, -0.83, 1.33, -0.02, 5.67, 1.19)$. Note that this relationship only depends on the ratio of subhalo to host halo mass, and not on redshift or host halo mass, which is a good approximation given the considerable other uncertainties in modeling the properties of dusty infrared sources within halos. For each dark matter host halo in the light cone catalog, we draw from this distribution in two steps. First, we determine the mean number of subhalos contained in a halo of a given mass,
\begin{equation}
\overline{N}_{\rm sh}(M) =\int_{M_{\rm min}}^M d\ln{m} \frac{dN_{\rm sh}}{d\ln{m}}.
\end{equation} 
Each halo is assumed to host one central infrared galaxy, following the infrared halo model discussed below in Section~\ref{subsubsec:cib-analytic}.  The number of satellites in a halo with mass $M$ is drawn from a Poisson distribution with mean of $\overline{N}_{\rm sh}(M)$, and its spatial distribution is  chosen with a  random angular position with respect to the halo center and with a radial distribution drawn from an NFW 
profile with a mass--redshift--concentration relationship given by \citep{2008MNRAS.390L..64D}. Once the number of satellites has been sampled from the Poisson distribution, each one is assigned a subhalo mass, $m$, drawn randomly from the subhalo mass function consistent with equation (\ref{eq:shmf}) and a minimum mass of $M_{\rm min}$.

%%%%%%%%%%%%%%%%%%%%%%%%%%%%
\subsubsection{Anisotropies Generated Exterior to Resolved Halos}
\label{subsubsec:gen-exterior}
In order to include Thomson scattering and lensing by material external to halos, we use a simplified model for the spatial distribution of matter outside of halos that uses Lagrangian perturbation theory. In particular, for each lattice site, $i$, in Lagrangian space that is outside of a halo at comoving position $\q_i=|\q_i|\nhat_i$, we determine the redshift corresponding to that point, $z_i=z(|\q_i|)$, assuming that a photon reaching the observer was emitted at that time from its Lagrangian redshift. As discussed in \S\ref{sec:lightcone}, the mass element belonging to that lattice site will have been displaced by peculiar motions, slightly changing the time at which the photon was emitted or scattered, but this is a negligible effect. Having determined the redshift corresponding to the mass element, we then determine its position as $\x_i=\overline{\q_i} + D(z_i)\sone_i + \Dtwo(z_i)\stwo_i$, where $\sone_i$ and $\stwo_i$ are the first and second order LPT displacements as determined in \S\ref{sec:ICs}. Once the perturbed position along the line of sight has been calculated, the total contribution from a given lattice site, $\delta{F}^{\rm f}_{ij}$, is determined by a redshift dependent kernel, such that 
\begin{equation}
\label{eq:field}
\delta{F}^{\rm f}_{ij}=\frac{a_{\rm latt}^3}{\Omega_{\rm pix}}W_F(z_i,\v_i\cdot\hat{\q})\left[
1+(b_F(z_i)-1)D(z_i)\delta^{(1)}(\q)
\right],
\end{equation}
where $a_{\rm latt}^3$ is the volume corresponding to each mass element or cell, $\v_i=a_i[\dot{D}(z_i)\sone_i+\Dtwodot(z_i)\stwo_i]$ is the peculiar velocity of the mass element, and $\Omega_{\rm pix}$ is the solid angle subtended by a pixel in the map. The factor $W_F(z,v)$ encapsulates the redshift of the source function and any line of sight peculiar velocity effects, and we describe the specific functional forms that we adopt for $W_F(z)$ in \S\ref{subsubsec:astro-exterior}. We also include a generic Eulerian bias factor $b_F(z)$, suitable for modeling biased electron or dusty galaxy distributions, but have set $b_F=1$ in the lensing and kSZ maps presented here.

As lattice sites nearby the observer can subtend an angular extent larger than an individual pixel of the map, we calculate the angular size of each lattice site, and split the lattice site in three dimensions into the required number of sub-volumes (to a maximum of 5$^3$ for computational efficiency), such that the angular extent of a sub-volume becomes smaller than a pixel. The total contribution of the parent lattice site is then split evenly between the sub-volumes, and each sub-volume is independently added as a point to the map. This particle sub-division suppresses the shot-noise at small angular scales. 

% ---------  ---------  ---------  ---------  ---------  ---------  --------- 
\subsection{Astrophysical Models}
\label{subsec:astro-model}
Here we present the fiducial astrophysical models for determination of the frequency dependent CIB intensity, $I^{\rm cib}_\nu(\nhat)$, the thermal SZ Compton-$y$ parameter, $y(\nhat)$, the kinetic SZ temperature fluctuation, $\Delta{T}^{\rm ksz}(\nhat)$, and the CMB lensing convergence, $\kappa^{\rm cmb}(\nhat)$. These were determined from halo profile fits to cosmological simulations and empirical models from the literature. We note that, for example, the CIB has various halo models used in the literature \cite{2017AstL...43..644P, 2016A&A...594A..23P, 2014A&A...571A..30P, 2013ApJ...772...77V, 2012ApJ...757L..23B, 2012MNRAS.421.2832S}, but exploring the differences of these when applied to our full-sky simulations is beyond the scope of this work. 

%%%%%%%%%%%%%%%%%%%%%%%%%%%%
\subsubsection{Thermal and Kinetic Sunyaev-Zel'dovich Effects}
\label{subsubsec:tsz-analytic}
The thermal Sunyaev-Zel'dovich (tSZ) effect is determined by the Compton y-parameter, which is proportional to the optical depth-weighted ratio of electron thermal and rest-mass energies in a given direction:
\beq
y = \int{d{\tau}} \frac{k_{\rm B}T_e(\chi \hat{n})}{m_ec^2} = \frac{8-5{\rm Y}_{\rm p}}{2(2-{\rm Y}_{\rm p})}\frac{k_{\rm B}\sigma_{\rm T}}{m_ec^2}\int d\chi (1+z)^{-1}P_{\rm th}(\chi \hat{n}),
\eeq
where $P_{\rm th}$ is the thermal pressure of the gas, $Y_{\rm p}\simeq 0.24$ is the abundance of helium, and the medium is assumed to consist of fully ionized hydrogen and helium. This expression is valid up to small relativistic corrections when the electron temperature is much greater than the background radiation temperature, $T_e{\gg}T_\gamma$. 
The spectral distortion induced has a characteristic effective temperature deviation given by \cite{1969Ap&SS...4..301Z}
\beq
\frac{\Delta{T_\nu}}{T}=\frac{d\ln{T}}{d\ln{I_\nu}}\frac{\Delta{I_\nu}}{I_\nu}=
\frac{1-e^{-x}}{x}\frac{\Delta{I}_\nu}{I_\nu}=
\left\{
\frac{x}{\tanh(x/2)}-4
\right\}y\equiv g(\nu)y,
\eeq
where $x\equiv h\nu/k_{\rm B}T$.

By using detailed hydrodynamical simulations including AGN feedback, \cite{2012ApJ...758...75B} determined a parametric model for the pressure profile over a broader range of redshift and halo mass than is accessible by current observations. Their profile is in good agreement with the results from X-ray observations of nearby galaxy clusters \cite{2010A&A...517A..92A}, Bolocam observations of massive galaxy clusters \cite{2013ApJ...768..177S}, those from the Planck Collaboration \cite{2013A&A...550A.131P}, and stacked SZ profiles on locally brightest galaxies \cite{2015ApJ...808..151G}. The dimensionless pressure profile is parameterized with the generalized NFW form given in equation (\ref{eq:gnfw}), such that $P_{th}(x|M,z) = P_\Delta(M,z)f(x|M,z)$, where $P_\Delta \equiv GM\Delta\rho_{c}(z)f_b/(2R_\Delta)$, which follows the self-similar scaling for galaxy clusters \cite{1986MNRAS.222..323K}. The overall amplitude amplitude, $F_0$, core-scale, $x_c$, and large radius asymptotic power-law index, $\beta$, are fit for as functions of $M$, and $z$, while the other power-law parameters, $\alpha$ and $\gamma$, are fixed are fixed to 1 and -0.3, respectively. The pressure profile parameters are obtained from the values in Table 1 of \cite{2012ApJ...758...75B}, corresponding to AGN Feedback with $\Delta=200$. 

Doppler shifting of CMB photons scattered by free electrons along the line of sight results in a black body temperature fluctuation given by 
\beq \label{eq:lensingT}
\frac{\Delta{T}^{\rm ksz}(\hat{n})}{T_{\rm cmb}}= -\int d\tau\ \v(\chi \hat{n})\cdot\hat{n}=
-\sigma_T\int d\chi (1+z)^2n_e(\chi\nhat)\v(\chi \hat{n})\cdot\hat{n},
\eeq
where $\tau$ is the Thomson scattering optical depth and $n_e$ is the comoving electron number density. The kSZ effect in our maps is separated into two components, a dense spherical halo and a clustered field, which includes electrons in the IGM, such that $\Delta{T}^{\rm ksz}(\nhat)=\Delta{T}^{\rm ksz}_{\rm halo}(\nhat)+\Delta{T}^{\rm ksz}_{\rm field}(\nhat)$. To prevent double counting of collapsed matter in the field, we compensate the halo map by subtracting out the kSZ effect from a uniform sphere with the same position and mass as the halo, but with a radius corresponding to an overdensity of $3$, somewhat smaller than than the Lagrangian radius of the halo, ensuring that mass and momentum is conserved on large scales.

We use halo gas density profiles fit to the same set of AGN simulations, with parameters given by \cite{2016JCAP...08..058B}. The dimensionless gas density profile is also parameterized with the generalized NFW form given in equation (\ref{eq:gnfw}), such that 
\begin{equation}
    n_e(x|M,z) = [1-Y_p/2]\overline{\rho}_b/m_pF(x|M,z),
\end{equation} 
where $\overline{\rho}_b$ is the mean comoving baryon density, and $Y_p=0.24$ is the helium mass fraction. Assuming all hydrogen and helium in gas is fully ionized, then $F$ corresponds to the gas overdensity. Similar to the pressure profiles, the overall amplitude amplitude, $F_0$, core-scale, $x_c$,  and large radius asymptotic power-law index, $\beta$, are fit for as functions of $M$, and $z$, while the other power-law parameters, $\alpha$ and $\gamma$, are fixed are fixed to 1 and -0.3, respectively. See \S\ref{subsubsec:astro-exterior} for the treatment of kSZ for the field component in material exterior to halos. 

%%%%%%%%%%%%%%%%%%%%%%%%%%%%
\subsubsection{Cosmic Infrared Background}
\label{subsubsec:cib-analytic}
The CIB is produced by star-forming galaxies when stellar radiation is absorbed by dust grains and re-emitted in the infrared. Star formation is dependent on host halo mass, environment, and redshift, and is suppressed at low and high masses by supernovae, active galactic nuclei (AGN), and other kinds of feedback. Since the clustering of halos and the galaxies within them source the observed intensity fluctuations, the CIB provides important empirical constraints on the connection between star formation and redshift over wide range of halo masses and redshifts, in particular for the faintest and most difficult galaxies to detect and study individually.

For this work we use the CIB halo model developed by \cite{2012MNRAS.421.2832S} and used by \cite{2014A&A...571A..30P} and \cite{2013ApJ...772...77V}. In particular we adopt the parameters used by \cite{2013ApJ...772...77V} to fit CIB power spectrum measurements with Herschel. We present the main details of the CIB halo model here, in addition to the elements that are unique to creating mock full-sky observations, but refer the reader to the literature for additional details and discussion \cite{2012MNRAS.421.2832S, 2013ApJ...772...77V, 2014A&A...571A..30P}. In this model, the rest-frame SED of a given source depends on the frequency $\nu$ of observation, (sub)halo mass $M$, and the redshift $z$:

\beq
L_{(1+z)\nu}(M,z) = L_0 \Phi(z) \Sigma(M,z) \Theta[(1+z)\nu,  T_d(z)],
\eeq
where:
\begin{itemize}
    \item The spectral energy distribution, $\Theta[\nu, T_d]$, is a greybody at low frequencies and a power law at high frequencies,
\begin{equation}\label{eq:theta}
\Theta(\nu, z) \propto \begin{Bmatrix}
\nu ^{ \beta  }B_{ \nu  }(T_{ d }(z))  \quad  \nu < \nu_{ 0 };\\
\quad \quad \,  \nu ^{-\gamma} \quad  \quad \, \, \nu \ge \nu_{ 0 }.
\end{Bmatrix}
\end{equation}
Here $B_\nu$ denotes the Planck function, and $\beta=1.6$ is dependent on the physical nature of the dust. The relative normalization of the two components (since the absolute normalization is already accounted for in $L_0$) are obtained by the requirement that the logarithmic slope is continuous at $\nu_0$, $\mathrm{dln}\Theta(\nu,z)/\mathrm{dln}\nu = - \gamma$. Finally, the redshift dependencies of the effective dust temperature is given by $T_d \equiv T_0(1+z)^{\alpha}$, where $T_0=20.7$ and $\alpha=0.2$. For the simulations used in this work, $z<4.6$, and the frequencies considered, $\nu<1000$, the spectral energy distribution remains a greybody. 

\item The redshift dependent global normalization of the $L-M$ relation is of the form 
\beq
\Phi(z) = (1+z)^{\delta_{CIB}},
\eeq
where $\delta_{CIB}=2.4$.

\item A log-normal function $\Sigma(M,z)$ is used for the dependence of the galaxy luminosity on halo mass,
\beq
\Sigma(M,z) = \frac{M}{ (2 \pi \sigma^2_{L/M})^{1/2} } \exp\left[{-\frac{( {\rm{log_{10}}}\ M-{\rm{log_{10}}}\ M_{eff})^2}{2 \sigma^2_{L/M}}} \right].
\eeq
$M_{eff}$ here describes the peak of the specific IR emissivity, and $\sigma_{L/M}$ describes the range of halo masses which produce the luminosity. We use the values of $\rm{log}(M_{eff}/M_\odot)=12.3$ and $\sigma^2_{L/M}=0.3$ from model 1 of \cite{2013ApJ...772...77V}. These act to produce a peak in the $L-M$ relation describing the maximum average infrared emissivity per unit mass, due to the suppression at both the high and low mass end. 
\item $L_0$ is a free normalization parameter, which we chose to reproduce the Planck 2013 \cite{2014A&A...571A..30P} power spectrum results at 545 GHz for $\ell = 500$ in \S\ref{subsubsec:cib-results}.
\end{itemize}

Having determined a population of subhalos for each host halo in the catalog, we populate each subhalo of mass $m$ with a satellite galaxy with luminosity $L_\nu(m,z)$, where $z$ is the redshift of the parent halo. We also include the emission from a central galaxy for each halo of mass $M$ at redshift $z$, and assign it a luminosity $L_\nu(M,z)$. This treatment considers the $L-M$ relation of satellites and centrals to be equivalent and ensures that every halo contains at least one infrared source. 

%%%%%%%%%%%%%%%%%%%%%%%%%%%%
\subsubsection{CMB Lensing}
\label{subsubsec:cmb-lensing}
As light propagates through the large scale structure of the universe it is deflected by the gravitational attraction of mass along the line of sight. In generating our maps we use the Born approximation, in which the displacement of light rays from their original position on the sky is small. We note that for upcoming surveys, post-Born terms will likely be non-negligible \cite{2018PhRvD..98d3512B,2019JCAP...10..057F}; in principle, a detailed ray-tracing approach to CMB lensing that does not implement the Born approximation could be applied to our simulations, but we leave implementation of this to future work. Under the Born approximation, the effect of lensing on the primary CMB is completely specified by the convergence, $\kappa$, and we need only integrate the matter overdensity along the line of sight, weighted by the lensing kernel $W_\kappa(\chi)$,
\beq \label{eq:lensing}
\kappa(\hat{n}) = \int_0^{\chi_*} d\chi W_\kappa (\chi) \delta(\chi \hat{n}),
\eeq
where $W_\kappa (\chi) =  3\Omega_M H_0^2 (1+z) \chi (1-\chi/\chi_*)/2$
and $\chi_*$ is the distance to the source. In this case, we choose the CMB as our source, located at a distance $\chi_* \simeq 13.8\ {\rm Gpc}$. The projection of matter density to generate convergence maps is analogous to the kSZ, except the latter only uses electron density, and is weighted by the velocity. Given the considerable astrophysical uncertainty in the shape of the matter density profile in the presence of feedback and star formation in the group and cluster mass halos resolved by CMB observations, the matter in all halos is assumed to follow an NFW profile with $c\equiv r_{200}/r_s=7$, independent of mass and redshift, where $r_{200}$ is the radius enclosing 200 times the mean matter density. To account approximately for nonlinear infall in the outskirts of the halo, we extrapolate the profile using $\rho(r) = \rho_{NFW}(r_{200})(r/r_{200})^{-2}$ for $r_{200}<r<2r_{200}$.  
Additionally, halos whose virial radii subtend a solid angle less than twice that of a pixel in the map are considered unresolved and are included in the field component, along with all the other matter exterior to resolved halos. The distribution of field matter is determined as described in the next section. To prevent double counting, we follow a similar procedure that for the electron density in the kSZ calculation described in \ref{subsubsec:tsz-analytic}. This ensures that overall mass is conserved and the resulting convergence maps naturally obey $\langle\kappa\rangle=0$. 

Material beyond the maximum redshift of the simulation ($z=4.5$) is not resolved by either the halo or the field component, and is instead included by generating an uncorrelated gaussian random map with a power spectrum determined using the Limber approximation,
\beq
C_\ell^{\kappa\kappa}=\int_{4.5}^{z_*} dz\frac{d\chi}{dz}\frac{W^2_\kappa(z)}{\chi^2}P_{\rm m}(k=\ell/\chi,z),
\label{eq:kappa-limber}
\eeq
and adding the $z>4.5$ gaussian random  map to the $z<4.5$ halo+field map. We use the CAMB `takahashi' Halofit model \cite{2003MNRAS.341.1311S,2012ApJ...761..152T} with the parameter `lens\_potential\_accuracy' set to 2 in order to calculate $P_{\rm m}(k=\ell/\chi,z)$.

We perform lensing of the CMB itself using the \texttt{pixell} package \footnote{\url{https://github.com/simonsobs/pixell}}.  We first generate a Gaussian random CMB map of both temperature and polarization.  We reproject the Websky convergence map in the Healpix pixelization to a lensing potential map in the cylindrical projection and pixellization used in the \texttt{pixell} package. We then perform the lensing, at 1' resolution.  The statistics of the resulting lensed CMB are described in Sec.~\ref{subsubsec:lensing}.

\subsubsection{Scattering, and Lensing from Outside of Resolved Halos} \
\label{subsubsec:astro-exterior}
As mentioned in \S\ref{subsubsec:gen-exterior}, observables can originate from regions exterior to resolved halos. In particular, angular fluctuations on the largest scales in kinetic SZ and lensing are mostly determined by material outside of the group and cluster mass halos resolved in the simulations presented in this paper. Here we detail the specific expressions for going from the field matter distribution described in \S\ref{subsubsec:gen-exterior} to an observable map. In all cases, the effects are encapsulated by the field redshift-volume kernel function $W_f(z)$ defined by equation (\ref{eq:field}). These are:
\begin{eqnarray}
W_{\rm f}^{\tau}(z,v) &\equiv& f_e\frac{\rho_{\rm b,0}\sigma_T}{\mu_em_p}\frac{(1+z)^2}{\chi^2(z)}\\
W_{\rm f}^{\rm ksz}(z,v)&=&-\frac{v}{c}W_{\rm f}^{\tau}(z,v)\\
W_{\rm f}^{\rm \kappa}(z,v) &\equiv& \frac{3}{2}\Omega_M H_0^2 \frac{(1+z)\left[1-\chi(z)/\chi_*\right]}{\chi(z)},
% W_{\rm f}^{\nu}(z,v) &\equiv& \frac{\epsilon^{\rm dg}_\nu(z)}{\chi^2(z)} 
\end{eqnarray}
where $f_e$ accounts for the mean fraction of gas that is ionized with mean molecular weight of $\mu_e$. We account for the measurements of the cosmic mean electron fraction from hydrodynamical simulations in \cite{2018ApJ...853..121P} in an approximate way by setting $f_e = 0.9$.  We assume hydrogen is fully ionized and helium is once ionized at $z>3$ and fully ionized at $z<3$ to set the mean molecular weight of ionized gas.
% The emissivity density of the CIB is determined analytically using
% \begin{equation}
% \epsilon_\nu^{\rm dg}(z) = \int_0^{M_{\rm res}} \frac{dn(M,z)}{dM} \left[ L_{\nu(1+z)}(M,z) + \int_0^M  L_{\nu(1+z)}(m,z)\frac{dn_{\rm sh}(m,M)}{dm} dm\right] dM
% \end{equation}
In this work, electron and matter fluctuations in the field are assumed to follow Lagrangian perturbation theory, so $b_{\tau}=b_{\rm ksz}=b_{\kappa}=1$.  Conversely, the halos hosting the CIB flux in our model are mostly resolved within our simulation, so we do not include a CIB field component.

% while for dusty galaxies we use a modified version the empirical fitting formula used by Planck, such that \begin{equation}
% b_{\rm dg}(z)={\rm min}\left[b_{\rm P}(z),
% b_{\rm low}(z)\right],
% \end{equation}
% where generally $b(z)=b_0+b_1z+b_2z$. The parameters for $b_{\rm P}(z)$ correspond to those from \citep{2014A&A...571A..30P}, $\left[b^{\rm P}_0, b^{\rm P}_1, b^{\rm P}_2\right]=\left[0.82, 0.34, 0.31\right]$, while those for the low redshift bias of the field, $b^{\rm low}(z)$, was determined by matching the total clustering in the combined halo and field maps to the measure Planck auto power spectra, $\left[b^{\rm P}_0, b^{\rm P}_1, b^{\rm P}_2\right]=\left[0.6,0,0.6\right]$. At $z\gtrsim 2$, the field bias is the same as that in Planck, while at $z=0$ it is a factor of $\sim 0.75$ lower. This is consistent with the high-redshift clustering being dominated by dusty galaxies in unresolved halos with $M\lesssim 10^{13}M_\odot$. 

% ---------  ---------  ---------  ---------  ---------  ---------  --------- 
\section{The Websky Extragalactic Sky Mocks}
\label{sec:skymocks}
\subsection{Large-Scale Structure Simulation}
\label{subsec:simulation}
The synthetic extra-galactic mocks presented in this work are constructed from a (15.4 Gpc)$^3$, 12,288$^3$ particle lightcone generated using the mass-Peak Patch method \cite{2019MNRAS.483.2236S, 1996ApJS..103....1B}, which we summarized in Section~\ref{sec:method}. This was accomplished by creating periodic initial conditions for a (7.7 Gpc)$^3$, 6,144$^3$ particle simulation, placing an observer at each of the eight corners the volume, performing the eight runs in serial, and stitching the results together to generate a single seamless halo catalog over the full-sky. This `octant' method has the advantage of doubling the one dimensional resolution of the simulation, or equivalently raising the volume resolution by a factor of 8, while using the same memory footprint. The disadvantage is that it replicates structures on scales equivalent to the box size, but although a given region in the volume is replicated eight times in this configuration, it will in general be observed at a different time from a different direction, making each octant of the sky unique, while avoiding any discontinuities or artefacts. Any periodic effects are limited to very large angular scales, $\ell < 10$, and there is no repetition of structure along any given line of sight.

The cosmology used for the simulations is consistent with the results of Planck 2018 \cite{2018arXiv180706209P}: $\Omega_m$ = 0.31, $\Omega_\Lambda$ = 0.69, $\Omega_b$ = 0.049, h = 0.68, $\sigma_8$ = 0.81, and $n_s$ = 0.965, and the linear power spectrum used to create the initial density field was generated with CAMB\footnote{\url{https://camb.info/}}.  The run time of the simulation was only 3.84 hours on 1128 Intel ``Skylak'' 2.4 GHz cores of SciNet's Niagara cluster \cite{2019arXiv190713600P}, for a total runtime of 4336 hours and a peak memory footprint of 7.67 TB ($\sim 9$ floats per resolution element). 5.9TB of disk space was required to store the initial conditions for additional post-processing, and 33GB was required for the halo catalogue. The computational efficiency of the mass-Peak Patch method allows for many realizations of a simulation of this size, but this currently remains as future work. 

\subsection{Halo Catalogue}
%BEGIN FIGURE -------------
\begin{figure}
\begin{center}
\includegraphics[width=1.05\textwidth, trim = 40 0 0 0]{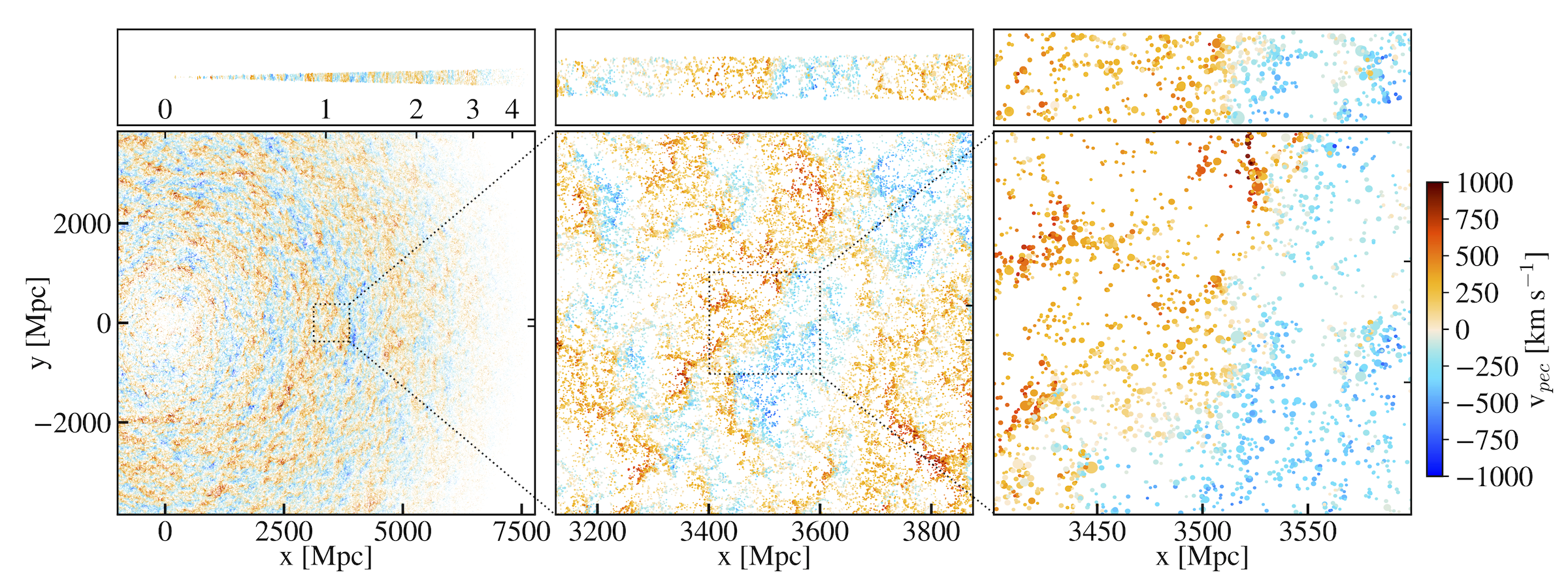}
\vspace{-0.8cm}
\caption{A thin equatorial wedge through the Websky lightcone, showing all halos within a degree of the equator. Top panels show the side view of a 1\degree\ beam, while bottom show the full top-down view of the equatorial wedge. The halo radius is proportional to the cube root of the halo mass, and colour depicts the peculiar velocity with respect to the observer. The redshift is indicated by the labels on top, while the comoving distance is indicated on bottom, and the ticks on the right axes of the bottom row indicate the angular extent of a 1\degree\ beam. The large-scale coherence of the velocity field on scales up to many hundreds of Mpc is immediately apparent, as well as the filaments, sheets, and voids that constitute the cosmic web.}
\label{fig:lightcone}
\end{center}
\end{figure}
%END FIGURE ----------------

We consider all halos within a distance $7.7{\rm Gpc}$ of the observer, corresponding to maximum redshift of $z= 4.6$ and a total survey volume of $\simeq 1,900$~$({\rm Gpc})^3$. Although the mass function of mass-Peak Patch halo catalogues is generally within $\sim 10\%$ of N-body \citep{2019MNRAS.483.2236S}, to ensure mass-matching with an equivalent N-body run we performed abundance matching along the lightcone to $M_{200\overline{\rho}_m}$ of \cite{2008ApJ...688..709T}. Abundance matching was performed by calculating the mass-Peak Patch mass function, $N(>M | z)$, in redshift bins of $\Delta z = 0.1$. For each halo mass bin, at each redshift, we determine the fractional change in mass needed for the number of mass-Peak Patch halos in the equivalent volume to match that predicted by \cite{2008ApJ...688..709T}, and save this in a two-dimensional table. The mass of each halo along the lightcone is then multiplied by a fractional change determined by bi-linear interpolation from this table. We retain halos in the catalogue with a pre-abundance matched mass greater than 10 particles of the simulation. This results in a redshift dependent mass completeness of the final catalogue of roughly $M_{\rm{min, Websky}} \simeq 1.2 \times 10^{12} M_\odot$ between $0< z < 4$ and increasing to $\simeq 4 \times 10^{12} M_\odot$ by $z=4.6$. An exact tabulation of the halo mass resolution as a function of redshift is provided alongside the halo catalogue.

The final catalogue contains roughly $9\times 10^{8}$ halos, each with an initial Lagrangian position [Mpc], final Eulerian position [Mpc], velocity [km/s], and mass [$M_{200\overline{\rho}_m}$], for a total of 10 floats per halo. The resulting catalogue and simulated maps are made publicly available as part of the Websky suite of extragalactic sky mocks\footnote{\url{http://mocks.cita.utoronto.ca/Websky}}. 

Shown in Figure \ref{fig:lightcone} is a thin equatorial wedge of the halo catalogue subtending a degree, representing slightly less than a percent of all the halos in this realization. The cosmic web of groups and clusters can be easily identified in the right two panels, with filaments, sheets and voids several tens of Mpc across, and smaller groups clustered around clusters. The peculiar velocity from the observer, illustrated by the halo's colour, shows the coherent large-scale velocity flows on scales much larger than the density, of up to many hundreds of comoving Mpc. The octant method used to generate the simulation assures no repetition of structure or velocity information along the line of sight, and large-volume simulations such as the one presented here are required to realize accurate large-scale velocity flows.

%%%%%%%%%%%%%%%%%%%%%%%%%%%%
\subsection{Extragalactic Maps}
\label{subsec:maps}

\begin{figure}[t]
\begin{center}
\includegraphics[width=1.0\textwidth, trim = 0 0 0 0]{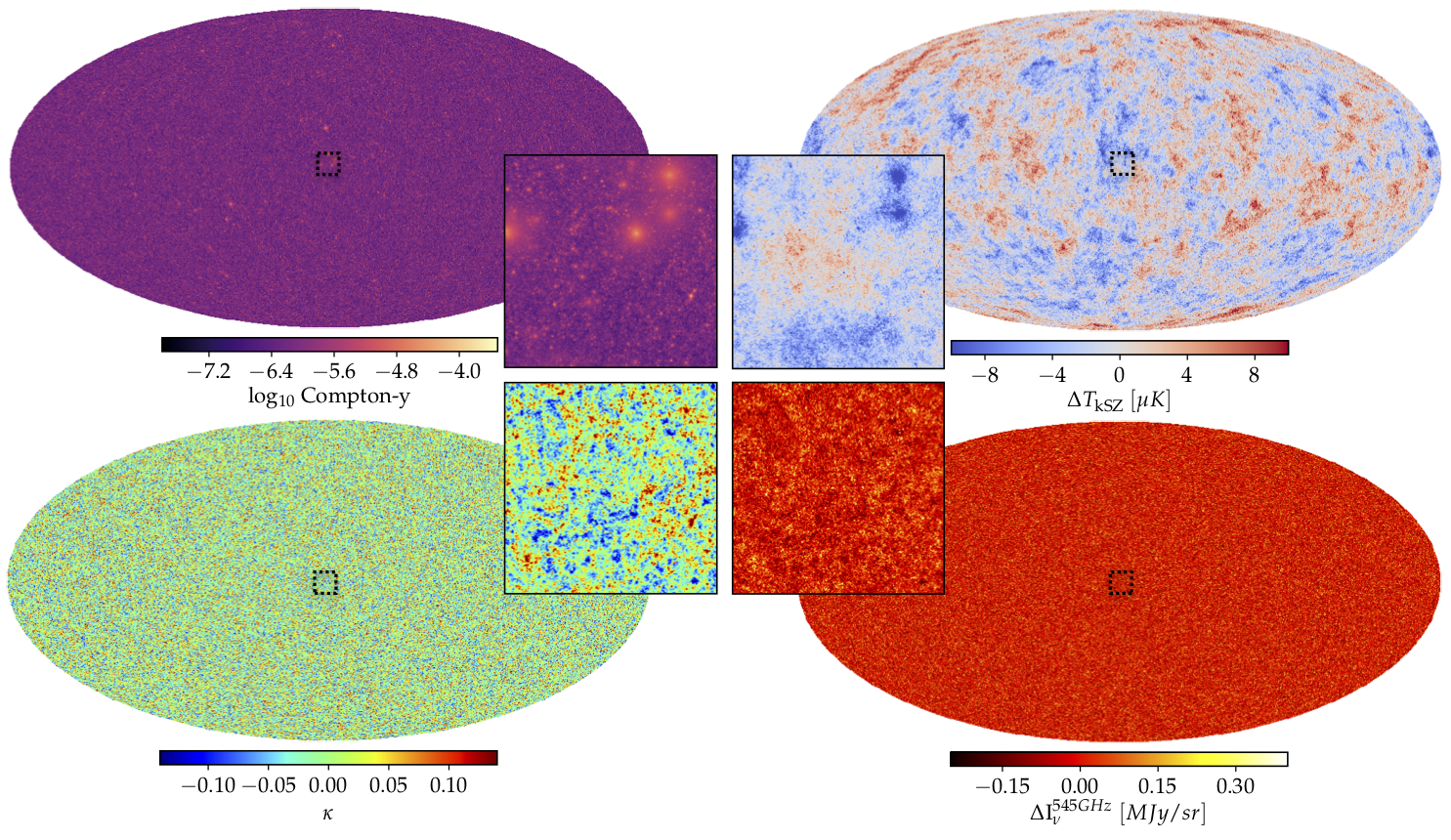}
\vspace{-0.8cm}
\caption{Clockwise from top-left: thermal Sunyaev-Zel'dovich, kinetic Sunyaev-Zel'dovich, Cosmic Infrared Background, and CMB lensing convergence signals from the Websky simulations. Central panels show a 10$\degree$ $
\times$ 10$\degree$  zoom in at the native resolution of the publicly available maps. The general correlation between all sky components is apparent, but the varying physics contributing to each effect, and different evolution with redshift, results in visually distinct maps.}
\label{fig:fullskies}
\end{center}
\end{figure}

Shown in Figure \ref{fig:fullskies} are the full-sky maps determined by projecting the halo catalogue and initial density field using the four models described in \S\ref{sec:mapmaking}. Clockwise from top left we show the thermal Sunyaev-Zel'dovich, kinetic Sunyaev-Zel'dovich, cosmic infrared background, and the CMB lensing convergence signals, with a 10$\degree$ $
\times$ 10$\degree$ zoom in at the native resolution of the publicly available maps ($N_{\rm side}=4096$) in the center. The correlation of different map components is clearly visible. Perhaps most prominent is the effect of nearby cluster-mass halos in the Compton-$y$ map, illustrating the strong bias towards massive objects at low redshift that contribute most of the signal due to their very high temperatures and late formation times. The kSZ map also reflects electron density inhomogeneities  associated with massive nearby clusters, on small and intermediate scales, with the overall effect arising from the coherent velocity flows that dominate the structure on much larger scales. Having a large simulation volume approaching the cosmic variance limit is essential for simulating the SZ sky, most importantly to sample the large scale velocity modes that contribute significantly to the kSZ effect and to probe the tail of the mass distribution of the rarest galaxy clusters that is most sensitively probed by the tSZ effect. The CIB and CMB lensing convergence, on the other hand, reflect structure from higher redshifts and lower halo masses, where both the infrared emission and number of galaxies per sky-area peak, and consequently those maps are more diffuse in appearance and appear more correlated with each other than they do with the tSZ or kSZ maps. Bright unresolved galaxies contribute a notable flux to individual pixels of the CIB map (white), while the CMB lensing convergence is more smooth on small scales due to the diffuse halo profiles and the unity bias of the matter density field. The CIB (convergence) map was smoothed using a 2 (4) arcminute beam for visualization purposes.
%BEGIN FIGURE -------------
%END FIGURE -------------

%BEGIN FIGURE -------------
\newcommand{\cmbsize}{0.45}

\begin{figure}[t]
\begin{center}
\includegraphics[width=\cmbsize\textwidth, trim = 30 10 5 5, clip = True]{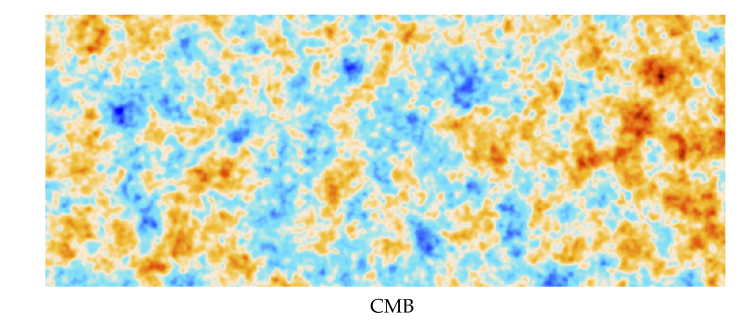}
\hspace{0.25cm}
\includegraphics[width=\cmbsize\textwidth, trim = 30 10 5 5, clip = True]{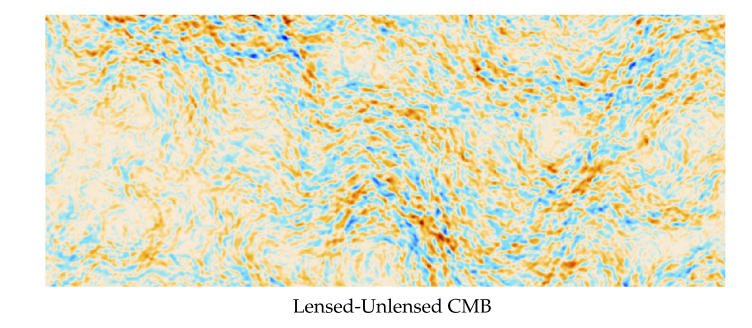}

\includegraphics[width=\cmbsize\textwidth, trim = 30 10 5 5, clip = True]{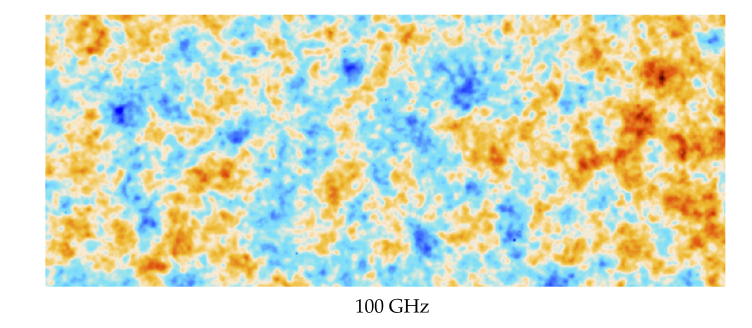}
\hspace{0.25cm}
\includegraphics[width=\cmbsize\textwidth, trim = 30 10 5 5, clip = True]{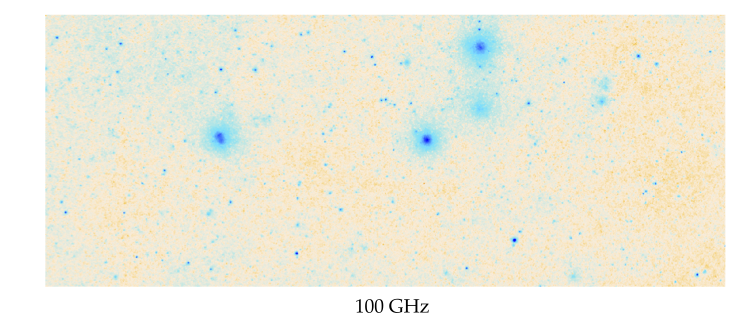}

\includegraphics[width=\cmbsize\textwidth, trim = 30 10 5 5, clip = True]{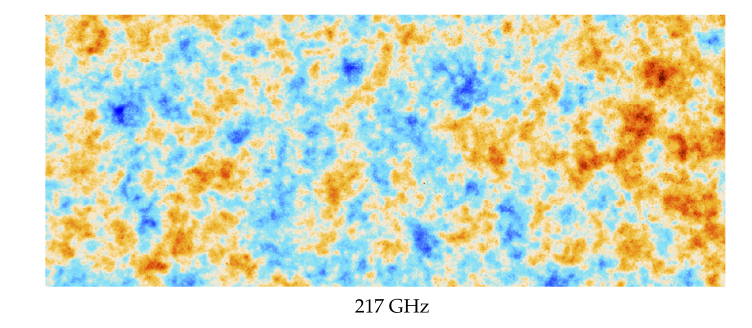}
\hspace{0.25cm}
\includegraphics[width=\cmbsize\textwidth, trim = 30 10 5 5, clip = True]{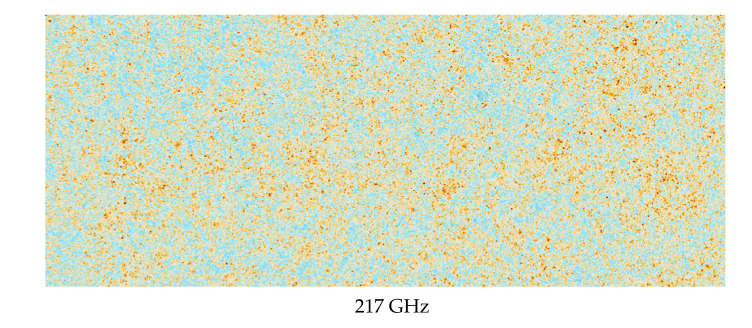}

\includegraphics[width=\cmbsize\textwidth, trim = 30 10 5 5, clip = True]{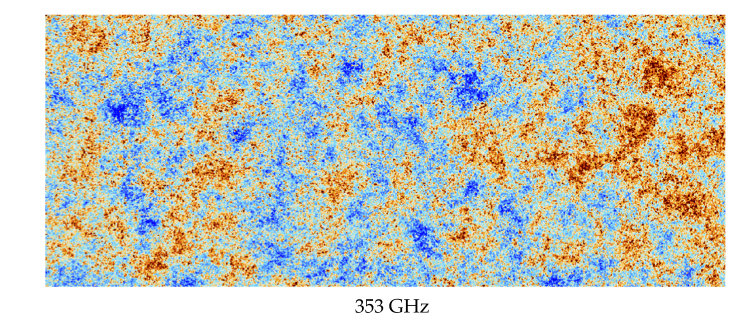}
\hspace{0.25cm}
\includegraphics[width=\cmbsize\textwidth, trim = 30 10 5 5, clip = True]{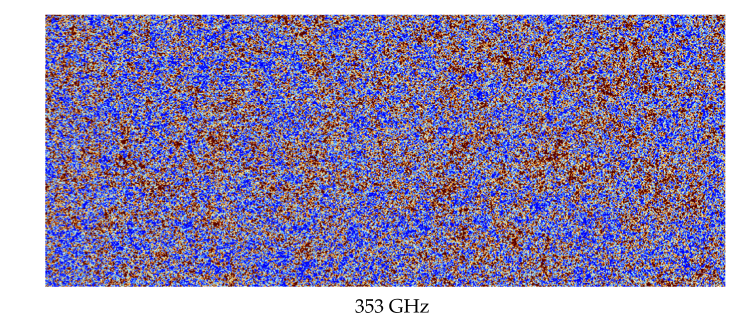}

\includegraphics[width=\cmbsize\textwidth, trim = 30 0 -5 5, clip = True]{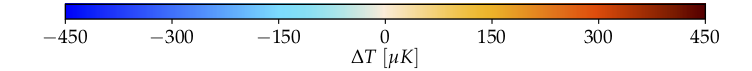}
\hspace{0.25cm}
\includegraphics[width=\cmbsize\textwidth, trim = 30 0 -5 5, clip = True]{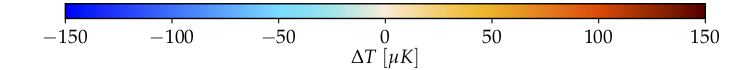}

\vspace{-0.7cm}
\end{center}
\caption{Mock observations of a 20$\degree$ $\times$ 8$\degree$ region of the CMB smoothed using a 1.4 arcminute beam. The top-left is CMB only, while the top-right shows the difference in lensed and unlensed CMB. The panels below on the left show total intensity fluctuations, including CMB, while the panels below on the right contain only the Websky CIB, tSZ, and kSZ contributions. The mean intensity has been subtracted, with intensity displayed in $\mu{K}_{\rm CMB}$ units (frequency-dependent conversion factors are given in Appendix~\ref{sec:unitappendix}). } 
\label{fig:cmb-maps}
\end{figure}
%END FIGURE ----------------

Using the Websky convergence map we lens a randomly generated primary CMB to create a lensed CMB map, as described in Section \ref{subsubsec:cmb-lensing}. This completes the set of extragalactic foreground maps presented in this work. To compare the extragalactic maps on an equal footing and illustrate their effects when added as foregrounds to the primordial CMB a few frequency dependent unit conversions are required, which we describe in Appendix~\ref{sec:unitappendix}.
        
Figure \ref{fig:cmb-maps} shows the full effects of the extragalactic foregrounds on observations of the primordial CMB at a range of Planck frequency channels from 100 GHz to 353 GHz. The mean temperature at each individual frequency has been subtracted, so that $\Delta T$ represents the fluctuations about the mean sky temperature at that frequency, not about $T_{\rm cmb}$. We show the primordial CMB in the top left panel, while the remaining panels on the left show the observed CMB after adding the four extragalactic foregrounds simulated in the Websky suite: tSZ, kSZ, CIB, and lensing. The top right panel shows the effect of lensing, while the other panels on the right hand side show the SZ and CIB foregrounds with no CMB contribution, using a different colour scale. Lensing was kept distinct to more easily distinguish the other foreground effects at low frequencies. The frequency dependent conversion factors for Compton-y and the CIB, coupled with the redshift evolving SED inherent in the CIB model, result in visually distinct effects in the observed CMB maps as a function of frequency. 

Below 217 GHz, lensing is the dominant large-scale foreground, introducing a characteristic hot/cold variation wherever positive/negative features in the lensing convergence map happen to align with gradients in the primordial CMB. The standard deviation of the lensing effect is $\sim$20 $\mu$ K, with a minimum and maximum value pixel value of roughly $\pm$ 200. Focusing beyond lensing in the remaining panels on the right hand side, we see that the CIB is the dominant foreground at frequencies above 100GHz, as expected. By 353 GHz the CIB has begun to add significant signal to the primary CMB, and by 545 and 857 GHz the primary CMB and all other foregrounds are nearly indistinguishable from the CIB contribution. For this reason we do not show frequency channels above 353 GHz. The 100GHz foreground map has a contribution from both the CIB and the the tSZ effect - which at this frequency shows up as a temperature decrement due to the upscattering of photos to higher frequencies when passing through clusters. At 217 GHz the tSZ has nearly no net effect, as the amount of photons upscattering out of the infinitesimally narrow band used here is equal to the number upscattering into the band from smaller frequencies, and the CIB again dominates. Below 217 GHz the tSZ is most apparent, and at 100 GHz the large scale flow of the kSZ is also slightly visible, alongside a subdominant kSZ contribution from individual large halos with relatively large peculiar velocities. 

%%%%%%%%%%%%%%%%%%%%%%%%%%%%
\subsection{Power Spectra}
\label{sec:powerspectra}
In this section we show the auto power spectra of thermal SZ, kinetic SZ, CMB lensing convergence, and the CIB at selected frequencies. We also show the cross-power spectra of the CIB with both thermal SZ and lensing, and the lensed TT, EE, and BB power spectra for a random primary CMB realization with the same cosmology. Where appropriate, we compare the power spectra to those from existing observations, simulations, and analytical predictions. 

\subsubsection{Thermal Sunyaev-Zel'dovich}
%BEGIN FIGURE -------------
\begin{figure}[t]
\begin{center}
\includegraphics[width=0.8\textwidth, trim = 0 0 0 0, clip = True]{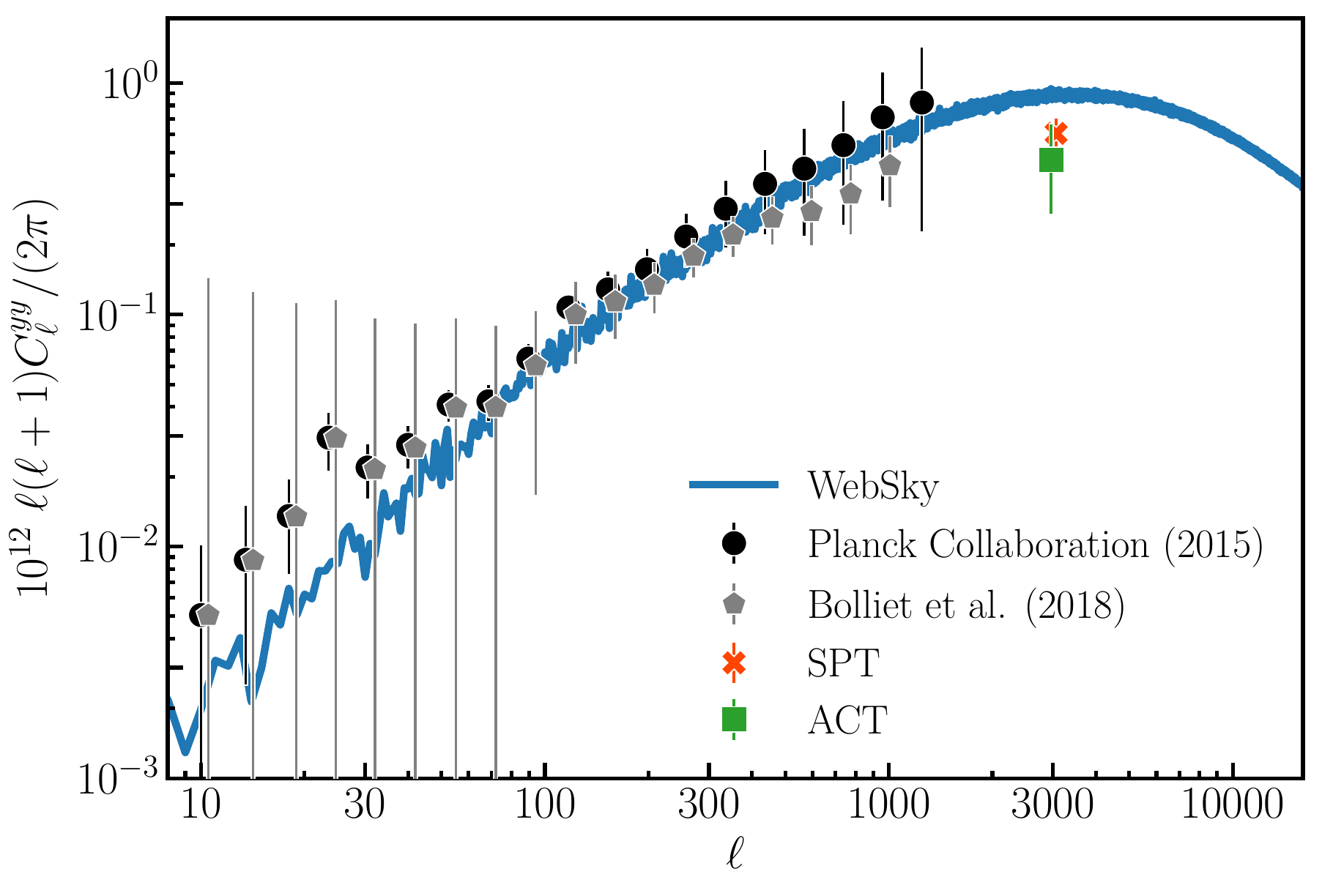}
\vspace{-0.7cm}
\end{center}
\caption{The power spectrum of the Websky Compton-y map (blue line). We also include the Planck measurement (black) \cite{2016A&A...594A..22P}, an external Planck re-analysis taking into account the trispectrum in the covariance matrix and an extended marginalization (gray) \cite{2018MNRAS.477.4957B}, the ACT measurement (green) \cite{2013JCAP...10..060S}, and the SPT measurement (orange) \cite{2015ApJ...799..177G}. }
\label{fig:tsz-power}
\end{figure}
%END FIGURE ----------------

% ACT TSZ a_tsz https://arxiv.org/pdf/1301.0824.pdf
%148 GHz is measured to be $D^{\nu=148GHz}_{\ell=3000} = 3.4 $\pm$ 1.4 µK^2$ \citep{2013JCAP...10..060S}
% PLANCK TSZ http://planck.caltech.edu/pub/2015results/Planck_2015_Results_XXII_Map_Thermal_SZ_Effect.pdf
% TSZ Planck reanalyzed https://arxiv.org/pdf/1712.00788.pdf
%\citep{2018MNRAS.477.4957B}
%SPT We measure the tSZ power at 143 GHz to be $D^{\nu=148GHz}_{\ell=3000} = 4.08^{+0.58}_{-0.67} \mu K^2$ \citep{2015ApJ...799..177G}

Following the thermal Sunyaev-Zel'dovich halo prescription described in \S\ref{subsubsec:tsz-analytic} we projected the pressure profiles determined from the hydrodynamical simulations of \cite{2012ApJ...758...75B} onto the mass-Peak Patch halo catalogue to create the Websky Compton-y map. All $\sim$9 $\times$ 10$^{8}$ halos were used, and no contribution from the field was considered. Although the minimum halo mass used to determine the pressure profiles was $\sim$1$\times$10$^{13}$M$_\odot$, the uncertainty introduced by using these profiles for less massive Websky halos is neglegible compared to the total signal, due to the approximately $M^{5/3}$ scaling of pressure. Figure~\ref{fig:tsz-power} shows the resulting power spectrum (blue line). For reference we include the Planck results (black) \cite{2016A&A...594A..22P}, and the external Planck re-analysis (gray) which took into account the trispectrum in the covariance matrix and an extended marginalization  \cite{2018MNRAS.477.4957B}. We also include the ACT measurement (green) of $D^{\nu=148GHz}_{\ell=3000} = 3.4 \pm 1.4 \mu K^2$ \cite{2013JCAP...10..060S}, and the SPT measurement (orange) of $D^{\nu=143GHz}_{\ell=3000} = 4.08^{+0.58}_{-0.67} \mu K^2$ \cite{2015ApJ...799..177G}. 

We find good agreement with the three datasets at all values of $\ell$. The large-scale tSZ power spectrum below $\ell \sim 300$ is highly dependent on the individual cosmological realization due to cosmic variance. In a similar study where we created $\sim 200$ full-sky realizations of a smaller volume to $z=1.25$ we found variance contours at $\ell < 100$ similar in size to those shown for \cite{2018MNRAS.477.4957B}, but the Websky realization aligns well with Planck at these large scales. The small scale power spectrum ($\ell >$ 10$^3$), on the other hand, is sensitive to the details of the halo pressure profiles used, as it directly encodes the shape of the gNFW profile, integrated over the redshift-evolving mass function and halo viewing angle. We find a slight increase in the power at $\ell$ = 3000 when compared to both the SPT and ACT measurements. Whether the disagreement is due to systematic effects in the SPT and ACT measurements, or indicates a physically relevant small-scale suppression in the tSZ power is not clear, due to the limited frequency coverage and sensitivity of SPT and ACT.

A free parameter when pasting on diffuse profiles is the integration radius at which to cutoff the halo profile. Due to the small radius power-law index of $\gamma = -0.35$, the pressure profile in the inner regions of halos falls more slowly with radius compared to the density profile, while beyond a radius of $\sim R_\Delta$ the signal quickly falls due to large radius power-law index of $\beta \sim -4.35$ (with a slight redshift dependence). Therefore, only integrating profiles to a small radius results in not accounting for signal from the outskirts of halos, and too large of radius will cause excessive double-counting of signal from neighbouring halos. For example, when using $R_{max}=[2,3,4]R_\Delta$, the total Compton-y contribution of a $1\times10^{14}M_\odot$ halo at redshift zero is a factor of $\sim[1.9, 2.4, 2.7]$ larger than if stopping the integration at $R_{max}=R_\Delta$. In the power spectrum this effect mainly shows up as a slight normalization change for the range of multipoles shown, except for at the very smallest scales where the shape and extent of the profile is reflected. Since we do not attempt to model the thermal pressure distribution of material not associated with the resolved halos from the simulation, we chose an integration radius of $4R_\Delta$ for this study. If one was to use the resulting Compton-y map for a detailed investigation of e.g. filamentary structures, the effects of this choice resulting in potential double counting of signal in the outskirts of halos would first need to be studied in further detail beyond the power spectrum analysis performed here.    

\subsubsection{Kinetic Sunyaev-Zel'dovich}

The kinetic Sunyaev-Zel'dovich map contains both a spherical halo profile, and a clustered field component containing the electrons in the intergalactic medium and unresolved halos, described in \S\ref{subsubsec:tsz-analytic} and \S\ref{subsubsec:astro-exterior}, respectively. All halos for which $r_{\rm 200c}$ subtends greater than half an arcminutue and $M_{\rm 200c}$ is greater than 10$^{13}$M$_{\odot}$ were used in the creation of the halo map, and material in halos smaller than this was considered to be part of the field.  This distinction was implemented due to the minimum halo mass used to determine the free electron profiles of $M_{\rm 200c}\sim\times$10$^{13}$M$_\odot$.

\begin{figure}[t]
\begin{center}
\includegraphics[width=0.9\textwidth, trim = 0 0 0 0, clip = True]{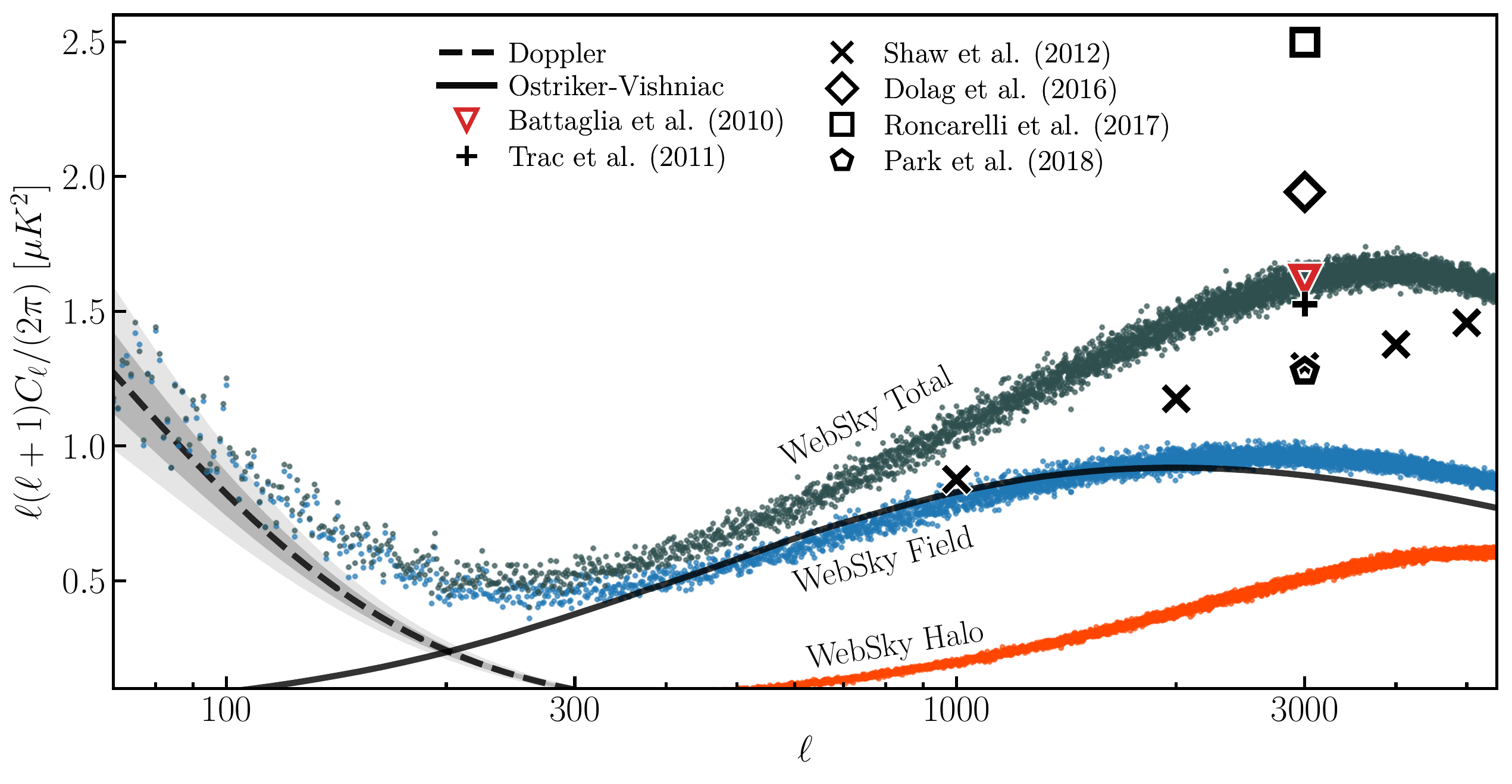}
\vspace{-0.7cm}
\end{center}
\caption{The power spectrum of the total Websky kSZ map (gray), and the individual contributions to the power from the halos (orange) and the field (blue). We also include the available high-resolution simulation results of Battaglia et al. (red triangle) \cite{2010ApJ...725...91B}, Trac et al. (black plus) \cite{2011ApJ...727...94T}, Shaw et al. (black X) \cite{2012ApJ...756...15S}, Dolag et al. (black diamond) \cite{2016MNRAS.463.1797D}, Roncarelli et al. (black square) \cite{2017MNRAS.467..985R}, and Park et al. (black pentagon) \cite{2018ApJ...853..121P}. We depict the Battaglia et al. result with a different colour as it was measured from the same set of hydrodynamical simulations used to fit the free electron profiles applied in this work. The data points for all non-Websky simulations are adapted from Table 1 of \cite{2018ApJ...853..121P}, where we first used the radiative cooling and star formation (CSF) scaling relation given in Table 3 of \cite{2012ApJ...756...15S} to scale their results to our cosmology, and assumed that the ratio of powers between models at $\ell=3000$ holds under the cosmological scaling performed. We also include analytical calculations of the Doppler contribution (dashed black) resulting from the sharp cutoff of structure at z=4.5 in the Websky simulations, and the contribution from Ostriker-Vishniac component (solid black) \cite{2016ApJ...824..118A}.}
\label{fig:ksz-power}
\end{figure}

The power spectra of the Websky halo (orange), field (blue), and total (gray) kSZ map are shown in Figure~\ref{fig:ksz-power}. For comparison we include available high-resolution simulation results including: the hydrodynamical simulations of Battaglia et al. (red triangle) \cite{2010ApJ...725...91B}, Shaw et al. (black X) \cite{2012ApJ...756...15S}, and Roncarelli et al. (black square) \cite{2017MNRAS.467..985R}, the post-processed dark matter simulations of Trac et al. (black plus) \cite{2011ApJ...727...94T}, 
the Magneticum results of Dolag et al. (black diamond) \cite{2016MNRAS.463.1797D}, 
and the Illustrius results of Park et al. (black pentagon) \cite{2018ApJ...853..121P}. We depict the Battaglia et al. result with a different colour as it was measured from the same set of hydrodynamical simulations used to fit the free electron profiles applied in this work. The data points for all non-Websky simulations are adapted from Table 1 of \cite{2018ApJ...853..121P}, where we first used the radiative cooling and star formation (CSF) scaling relation given in Table 3 of \cite{2012ApJ...756...15S} to scale their results to our cosmology, and assumed that the ratio of powers between models at $\ell=3000$ holds under the cosmological scaling performed. We find that the kSZ power spectrum is dominated at all scales by the contribution from the field component, with the halo component  contributing roughly 5/6ths of the total signal at $\ell = 5000$. The resulting total kSZ power spectrum is in rough agreement with the available simulation based predictions. 

%BEGIN FIGURE -------------
%END FIGURE ----------------

Given that the field component is based on second order Lagrangian perturbation theory, we expect the resulting power spectrum to be approximately consistent with the leading order contributions to the kSZ power spectrum from linear density and velocity fluctuations. The kSZ contribution in a given direction $\hat{n}$ can be written as
\beq
\frac{\Delta T_{kSZ}(\hat{n})}{T_{CMB}} = \int (1+\delta_e) \v \cdot \hat{n} d\overline{\tau} =  \int (1+\delta + \delta_x + \delta \delta_x) \v \cdot \hat{n} d\overline{\tau},
\eeq
where
\begin{equation}
\overline{\tau} = \int \sigT n_{{\rm e},0}x(z)(1+z)^2d\chi,
\end{equation}
$\delta_e$ is the electron density contrast, $\delta$ is the gas density contrast, $\delta_x$ is the ionization fraction contrast, $\v$ is the velocity, $x(z)$ is the mean volume averaged ionized fraction, and the remaining terms are described in \S \ref{subsubsec:tsz-analytic}. The first three contributions are commonly referred to as the `Doppler' ($\propto \v$), the 
`Ostriker-Vishniac' ($\propto \v \delta$), and the `patchy' ($\propto \v \delta_x$), respectively \cite{2016ApJ...824..118A}.
Although reionization is not included in these z<4.5 simulations, we find a large scale power of a few $\mu K^2$ resulting from the sharp cutoff in structure at the maximum redshift of the simulations, $z_{\rm max}$. This sharp cutoff suppresses cancellation of the line of sight velocities from comoving distances beyond the bounds of the simulation, with a power given by 
\begin{align}
%    C_\ell^D \equiv \int d\chi \int d\chi' \frac{\partial u_0}{\partial \chi} \frac{\partial u_0'}{\partial \chi'} W_\ell (\chi, \chi') 
%    &\approx \frac{1}{\ell^4}\int d\chi\chi^2\left(\frac{\partial u_0}{\partial \chi} \right)^2P(k=\ell/\chi) \\
%    C_\ell^{RD} \equiv u_0(z_*)\int d\chi \frac{\partial u_0}{\partial \chi}, W_\ell(\chi,\chi_*) 
%    &\approx \frac{u_0(z_*)}{\ell^4}\frac{\partial u_0}{\partial \chi}(z_*)\chi_*^2P(k=\ell/\chi_*) 
%    \\
    C_\ell^b = \frac{2f_e u_0^2(z_{\rm max})}{\pi} \int \frac{dk}{k^2} P(k) j_\ell(k \chi_{\rm max}) j_\ell(k \chi_{\rm max}), 
\end{align}
where $u_0(z) \equiv \sigma_Tn_{e,0}(1+z)Hf$ and $f\equiv d\ln{D}/d\ln{a}$. 
When integrating to a maximum redshift of $z_{\rm max}=4.5$, this is the dominant source of power at large angular scales. The Ostriker-Vishniac power spectrum dominates on small scales and is given by
\beq
C_\ell^{ov} \equiv \frac{f_e}{2\ell^2} \int \frac{d \chi}{\chi^2} u_0^2(z) \int \frac{d^3 \k'}{(2 \pi)^3} P(|\k - \k'|) P(k') \frac{k(k-2k'\mu)(1-\mu^2)}{k'^2(k^2 +k'^2 - 2k k' \mu)},
\eeq
where $\mu \equiv \hat{\k} \cdot \hat{\k}'$.
We include the Doppler and Ostriker-Vishniac components in Figure~\ref{fig:ksz-power}.  Previous N-body simulations of the kSZ sky at z<1 \cite{2016ApJ...823...98F} contained similar descriptions of the field (model I), halo (model II), and total (model III) as shown here. While the different redshift coverage, model for the kSZ signal of clusters, radial extent of the cluster profile integration, and cosmological parameters, make a direct comparison impossible, we find relatively good agreement. 

\subsubsection{Cosmic Infrared Background}
\label{subsubsec:cib-results}
%BEGIN FIGURE -------------
\begin{figure}[t]
\begin{center}
\includegraphics[width=0.8\textwidth, trim = 0 0 0 0, clip = True]{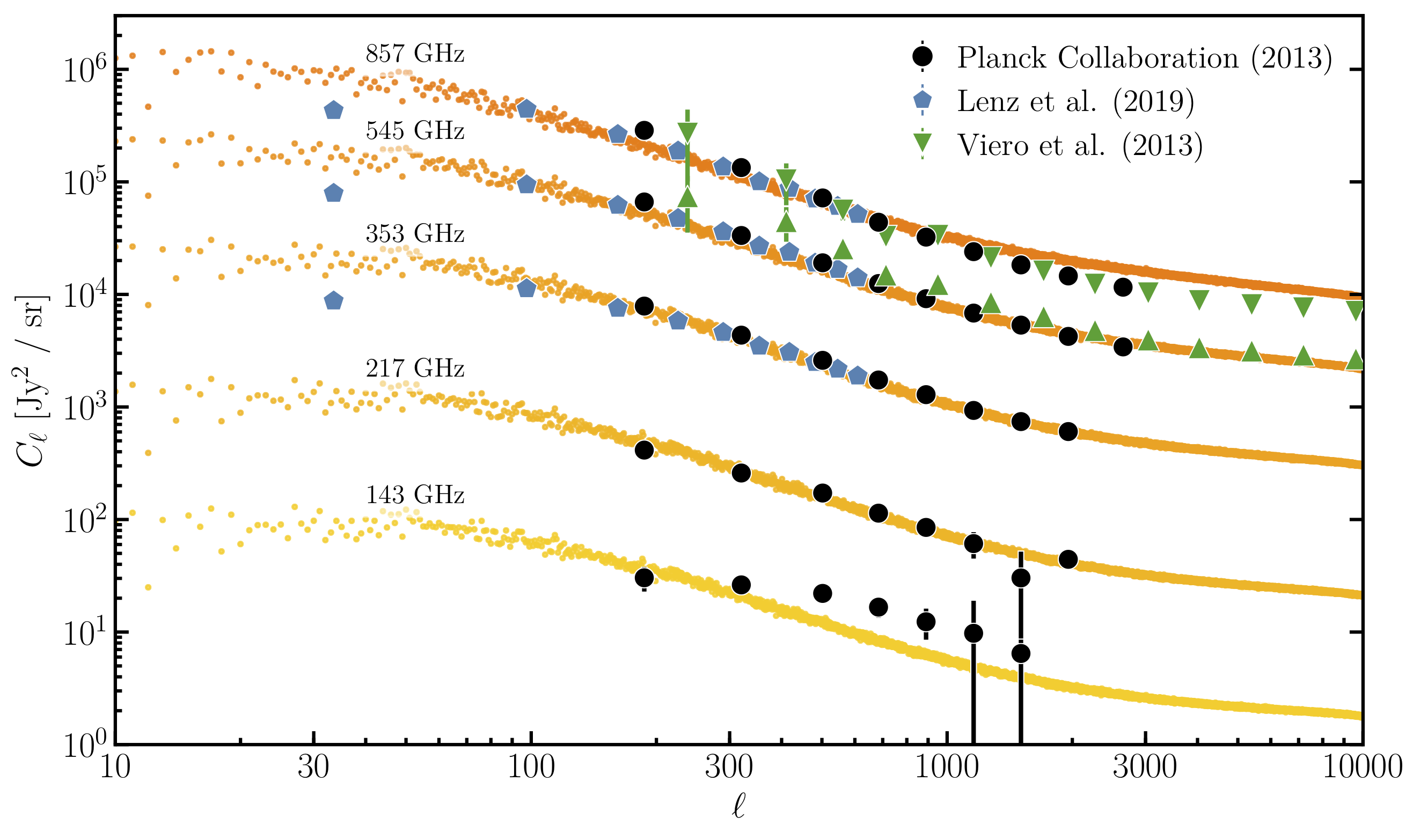}
\vspace{-0.7cm}
\end{center}
\caption{The power spectrum of the Websky Cosmic Infrared Background maps (orange) at the Planck HFI channels, and the corresponding Planck measurements (black circles) \cite{2014A&A...571A..30P}, the Herschel Multi-tiered Extragalactic Survey (HerMES) measurements of Viero et al. at 857 and 600 GHz (green triangles, downward and upward, respectively) \cite{2013ApJ...772...77V}, and the recent dust-cleaned Planck maps of Lenz et al. (gray pentagons) \cite{2019arXiv190500426L}. The Lenz et al. points are not shown at $\ell>700$ as they are qualitatively identical to Planck at small angular scales. The overall normalization factor $L_0$ required by the Planck CIB model used to construct the Websky maps was chosen to match the 545 GHz Planck results at $\ell=500$.}
\label{fig:cib-power}
\end{figure}
%END FIGURE ----------------

The Cosmic Infrared Background is simulated using the halo occupation distribution model described in \S\ref{subsubsec:cib-analytic}. We populated all $\sim$9 $\times$ 10$^{8}$ halos in the mass-Peak Patch catalogue with a number of central and satellite galaxies, resulting in a total of $\mathrm{ \sim3\times10^{9}}$ galaxies projected into the final map. Due to the mass resolution of the Websky halo catalogue of $\mathrm{\sim 1-2\times 10^{12}h^{-1} M_\odot}$, the CIB contribution from the field was not considered for this study. Maps were produced at the Planck HFI frequencies of 100, 143, 217, 353, 545, and 853 GHz, and a number of additional frequencies corresponding to the bands of current and future ground-based CMB experiments around this range. The Planck CIB model used requires the fitting of an overall normalization factor $L_0$, which we determined by requiring the power spectrum of the 545 GHz Websky map to match the 545 GHz CIB auto-power spectrum at $\ell=500$, as measured by the Planck Collaboration \cite{2014A&A...571A..30P}.

Figure~\ref{fig:cib-power} shows the power spectrum of the Websky CIB maps at the Planck frequencies with available data: 143, 217, 353, 545, and 853 GHz. We also include the 2013 results of the Planck collaboration (black circles) with the assumed radio-source shot-noise levels of $6.05 \pm 1.47$, $3.12 \pm 0.79$, $3.28 \pm 0.82$, $2.86 \pm 0.7$, and $4.28 \pm 0.90$ subtracted out \cite{2014A&A...571A..30P}, the Herschel Multi-tiered Extragalactic Survey (HerMES) measurements of Viero et al. (green triangles) \cite{2013ApJ...772...77V}, and the recent dust-cleaned Planck maps of Lenz et al. (gray pentagons) \cite{2019arXiv190500426L}. The HerMES results were obtained using the SPIRE instrument at 250, 350, and 500 $\mu m$ (1200, 857, and 600 GHz). Here we include the 857 GHz results as downward-facing triangles, and the 600 GHz results (note not 545 GHz) as upward-facing triangles, both using the `only extended sources masked' measurements provided. The Lenz et al. Planck analysis used additional data of neutral atomic hydrogen from the recently-released HI4PI Survey to create and remove template maps of Galactic dust, allowing for a measurements to extend to larger scales. We show the baseline results of Lenz et al., which used a low $\rm{H_I}$ column density mask corresponding to a threshold of $N_{\rm HI} = 2.5\times 10^{20}\ {\rm cm}^{-2}$. We do not show their results at $\ell>700$ as they are effectively the same as those from the Planck at small angular scales.

We find good agreement on the shape and amplitude of the CIB power spectrum at all frequencies, with a slight excess of power on the Poisson tail out to $\ell=10,000$ at 857 GHz. The Planck and Hermes data shown here do not have equal flux density cuts applied. HerMes `extended sources masked' corresponds to a frequency-independent flux density cut of 400 mJy, while Planck uses frequency dependent flux density cuts of [350, 225, 315, 350, 710] mJy for [143, 217, 353, 545, and 857] GHz, respectively. For the purpose of the CIB power spectrum plots shown in this paper, we applied a simplistic 400 mJy flux density cut to the nside=4096 Websky maps, by setting the value of any pixel with a flux density greater than 400 mJy to the mean of the remaining pixels below the flux density cut. This resulted in nulling [3, 109, 1229, 12772, 96011] pixels out of 200 million for the [143, 217, 353, 545, and 857] GHz maps. Additionally, the Websky data shown in this work uses the ideal case of an infinitely narrow passband centered at each frequency, while Planck and HerMes inherently include different spectral response functions integrated over the width of the passband and some assumption about the shape of the source spectrum. Although not included here, this can have a $\sim$30\% effect on the power spectrum between the two observations \cite{2013ApJ...772...77V}, and one can expect a similar effect on the Websky results when treated equivalently. 

\begin{table} 
\begin{center}
\begin{tabular}{ c c|c c c c c} 

 & & 857 & 545 & 353 & 217 & 143 \\
\hline
\multirow{3}*{857} & \textit{Websky} & 1 & 0.933 $\pm$ 0.017 & 0.882 $\pm$ 0.021 & 0.838 $\pm$ 0.026 & 0.802 $\pm$ 0.032  \\ 
 & \textit{Planck} & 1 & 0.949 $\pm$ 0.005 & 0.911 $\pm$ 0.003 & 0.85 $\pm$ 0.05  \\ 
 & \textit{Lenz et al.} & 1 & 0.96 $\pm$ 0.01 & 0.91 $\pm$ 0.01 & 0.85 $\pm$ 0.05  \\ 
 
 \addlinespace
\multirow{3}*{545} & \textit{Websky} & ... & 1 & 0.960 $\pm$ 0.014 & 0.935 $\pm$ 0.018 & 0.9077 $\pm$ 0.025 \\ 
  & \textit{Planck} & ... & 1 & 0.983 $\pm$ 0.007 & 0.90 $\pm$ 0.05  \\ 
  & \textit{Lenz et al.} & ... & 1 & 0.98 $\pm$ 0.01 &   \\ 
\addlinespace

\multirow{3}*{353} & \textit{Websky} & ... & ... & 1 & 0.968 $\pm$ 0.014 & 0.945 $\pm$ 0.021  \\ 
 & \textit{Planck} & ... & ... & 1 & 0.91 $\pm$ 0.05  \\
   & \textit{Lenz et al.} & ... & ... & 1  &   \\ 

\addlinespace
\multirow{1}*{217} & \textit{Websky} & ... & ... & ... & 1 & 0.960 $\pm$ 0.019  \\ 

\end{tabular}
\caption{Frequency decoherence of the CIB measured by averaging $C_\ell^{\nu \nu'}/(C_\ell^{\nu \nu} C_\ell^{\nu' \nu'})^{1/2}$ over the range $150 < \ell < 1000$. Error bars correspond to the standard deviation in this range. We include the Planck measurements of \cite{2014A&A...571A..30P} and the Lenz et al. measurements of \cite{2019arXiv190500426L}.}
\label{tab:cib_crosscorr}
\end{center}
\end{table}

As well as the powersectrum measurements, we determine the frequency decoherence of the Websky CIB maps through the cross correlation $C_\ell^{\nu \nu'}/(C_\ell^{\nu \nu} C_\ell^{\nu' \nu'})^{1/2}$. The CIB is expected to decohere between channels due to the redshift-dependent peak in the CIB distribution. We report this quantity averaged over the range $150 < \ell < 1000$ in Table~\ref{tab:cib_crosscorr}, where we restrict to this $\ell$ range to focus on the clustered CIB contribution and negate the contribution from shot noise as done in the Planck 2013 analysis. We neglect the decoherence with the 143 GHz map due to the high shot-noise contribution dominated by the radio sources \cite{2014A&A...571A..30P} which were not included in our study. We find similar results to the Planck measurements of \cite{2014A&A...571A..30P} and the Lenz et al. measurements of \cite{2019arXiv190500426L}, with significant decoherence between 857 and 217 GHz, and a small but non-zero decoherence for neighbouring frequencies.

\subsubsection{Weak Gravitational Lensing} \label{subsubsec:lensing}
%BEGIN FIGURE -------------
\begin{figure}[t]
\begin{center}
\includegraphics[width=0.7\textwidth, trim = 0 0 0 0, clip = True]{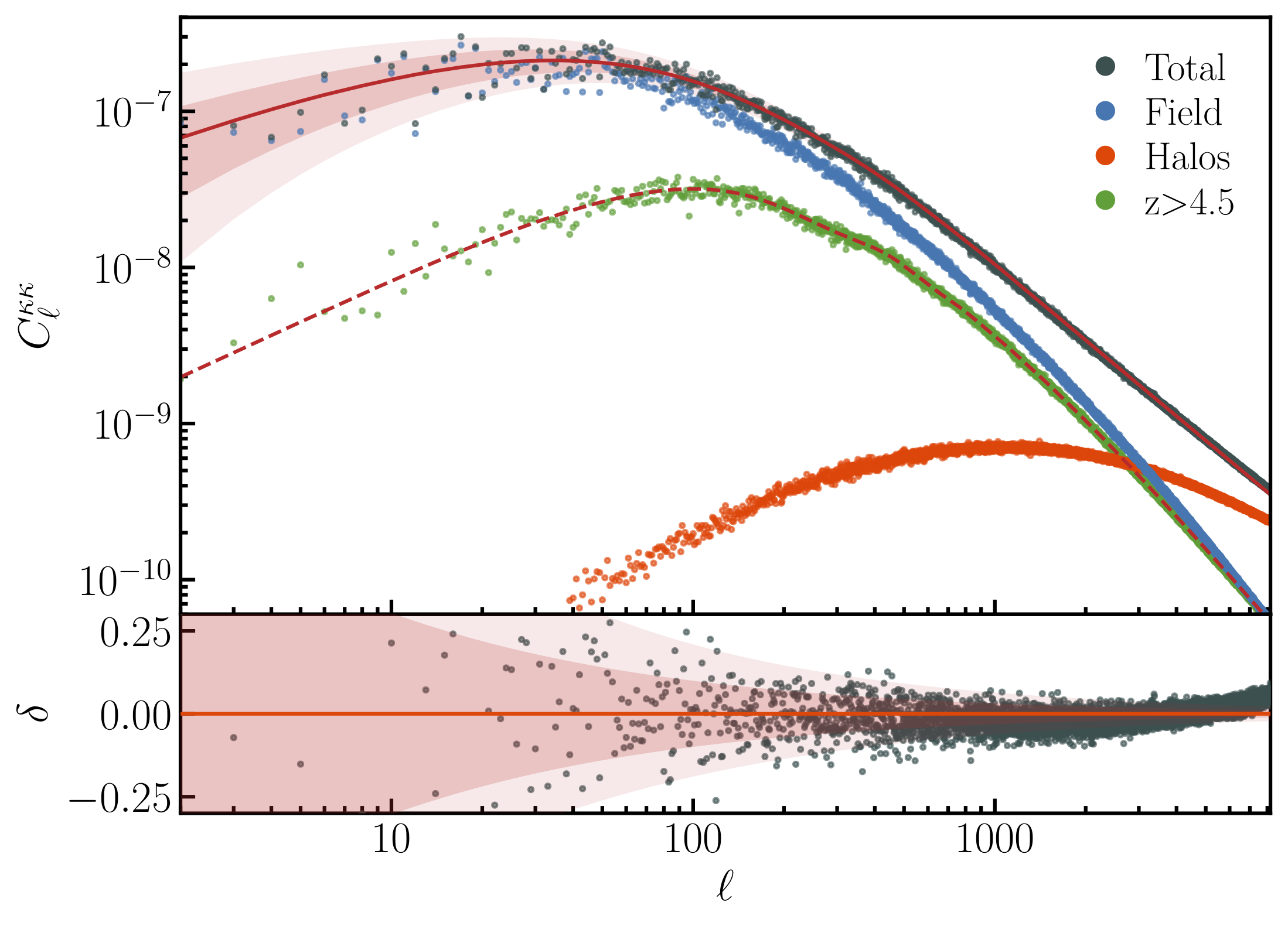}
\vspace{-0.7cm}
\end{center}
\caption{The lensing convergence power spectrum of the total Websky map (gray) and the Halofit model with cosmic variance errorbars (red). We also show the contribution from the three components necessary to construct the Websky map: halos (orange), field (blue), and the $z>4.5$ contribution from material beyond the redshift resolved in the dark matter simulations used (green).}
\label{fig:kappa-power}
\end{figure}
%END FIGURE ----------------

%BEGIN FIGURE -------------
\begin{figure}[t]
\begin{center}
\includegraphics[width=1.0\textwidth, trim = 0 0 0 0, clip = True]{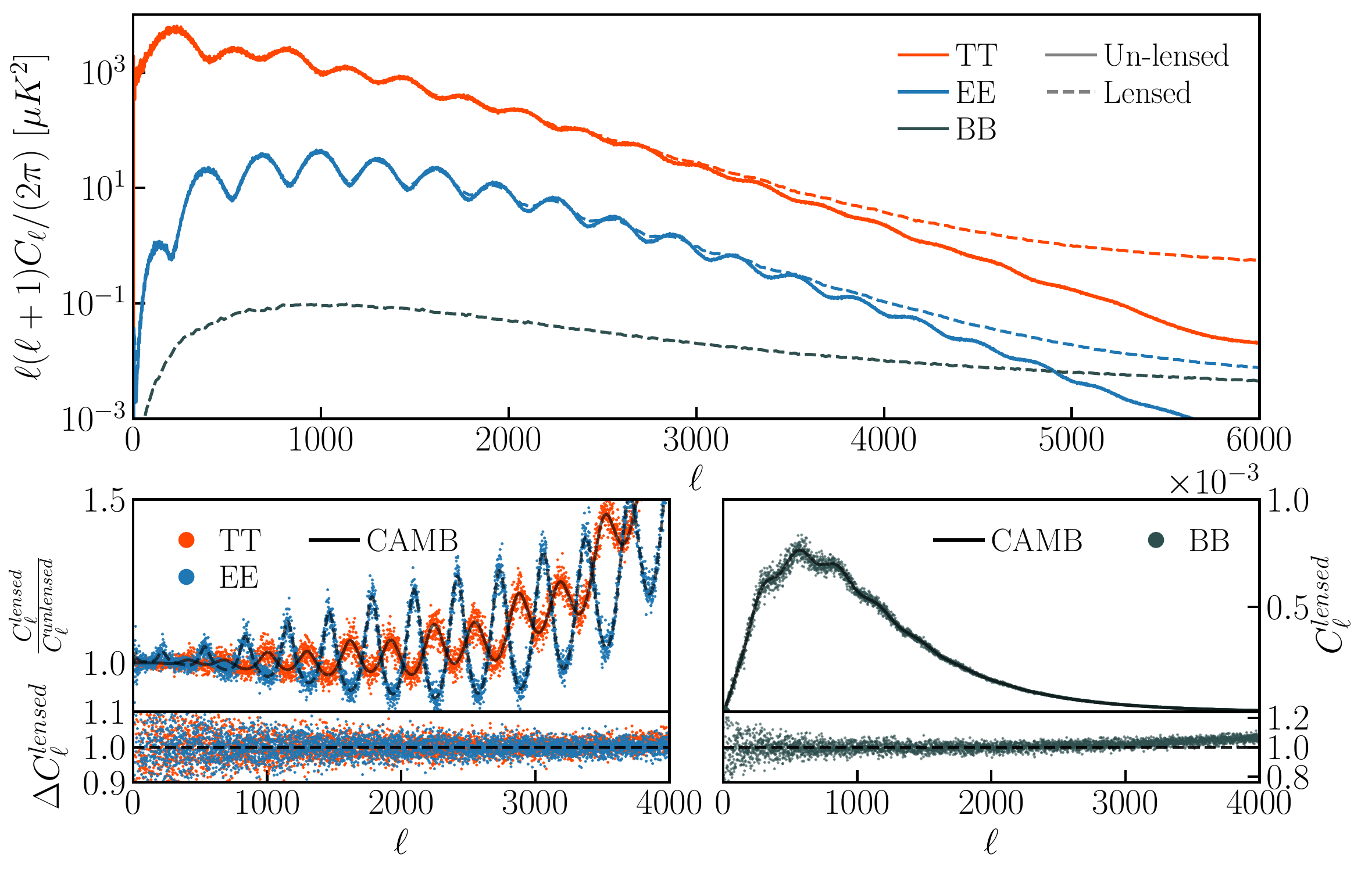}
\vspace{-0.7cm}
\end{center}
\caption{CMB lensing power spectra. Shown in the top panel are TT, EE and BB of unlensed (solid) and lensed (dashed) primary CMB temperature and polarization anisotropies, based on the maps generated with the procedure described in section \ref{subsubsec:cmb-lensing}. The unlensed BB power is zero since the unlensed primary maps were generated using a transfer function with tensor-to-scalar ratio $r=0$. Clearly evident is the well-known `peak-smearing' effect lensing has on the acoustic peaks, and the generation of lensing B-modes. The lower-left panel shows the ratio of lensed to unlensed power for TT and EE for our maps (dots), along with the prediction from CAMB (dashed and solid lines for EE and TT, respectively), with the same Halofit model for the matter power spectrum that was used in Figure \ref{fig:kappa-power}. The lower-right panel shows the same BB power from the lensed CMB map as shown in the top panel (dots), as well as the prediction from CAMB and Halofit. We find excellent agreement at all $\ell<5000$.}
\label{fig:lensing-power}
\end{figure}
%END FIGURE ----------------

The CMB lensing convergence map $\kappa$ and its lensing of the primary CMB are described in \ref{subsubsec:cmb-lensing}. The Websky CMB convergence map is constructed using the field particles and an NFW profile for all halos in the mass-Peak Patch halo catalogue that subtend more than a pixel, since lower mass halos are not resolved. We have added an additional component to the convergence, as an uncorrelated Gaussian random field, to account for the power originating from redshifts larger than the maximum included in the simulation, z$_{\rm{max}}$=4.5.  

Given the matter power spectrum as a function of redshift, $P_m(k,z)$ the angular power spectrum of the lensing convergence, $\kappa$, is well-approximated by the Limber approximation of Equation~\ref{eq:kappa-limber}. Shown in Figure \ref{fig:kappa-power} is the convergence found from Equation~\ref{eq:kappa-limber}, using the `Halofit' model for the matter power spectrum \cite{2003MNRAS.341.1311S,2012ApJ...761..152T} integrated from $z=0$ to $z\approx 1100$ (solid red), and integrated from $z=4.5$ to $z\approx 1100$ (dashed red). Also shown are the power spectra from the Websky halo (orange), field (blue), and gaussian $z>4.5$ (green) contributions, as well as the sum of the three (gray). The Websky and Halofit convergence power spectrum are consistent to within a few percent at $\ell < 1000$, with a residual suppression reaching $\sim 20$ percent at the smallest scales. This level of deviation from halo fit would be expected from 10 percent variations in halo density profiles that could be due, for example, to baryonic feedback effects. The large scale power is nearly entirely coming from material outside of resolved halos (field), while at $\ell > 3000$ the halo contribution to the power dominates. As expected, the field power spectrum follows closely that of the linear matter power spectrum on large scales.

Using the CMB convergence map, we lensed a randomly generated primary CMB following the procedure described in section \ref{subsubsec:cmb-lensing}, and show the power spectrum results of the simulations (top) and comparison to theory (bottom) in Figure~\ref{fig:lensing-power}. The simulation results for the temperature (TT) and  E-mode (EE) polarization show the well-known peak-smearing effect of lensing, where the acoustic peaks of the primary CMB (solid) become washed out into the lensed CMB (dashed), due to the local magnification and de-magnification effects of lensing which acts to transfer power between scales. The B-mode (BB) power of the unlensed CMB is identically zero due to the scalar-to-tensor ratio used as input, $r=0$, but the gravitational-lensing generated B-mode component is non-zero and matches well with theory, as seen in the bottom right panel. The bottom left panel shows the ratio of the lensed to unlensed power for both TT and EE for the simulations (coloured dots) and the theoretical calculation from CAMB (solid and dashed line for TT and EE, respectively), using the same Halofit model as for the kappa analysis discussed directly above. We find excellent agreement for TT, EE, and BB, when compared to the CAMB results for $\ell < 5000$. 

\subsubsection{CIB-Lensing Cross-correlation}

%%%%%%%%%%%%%%%%%%%%%%%%%%%%

%BEGIN FIGURE -------------
\begin{figure}[t]
\begin{center}
\includegraphics[width=0.98\textwidth, trim = 0 0 0 0, clip = True]{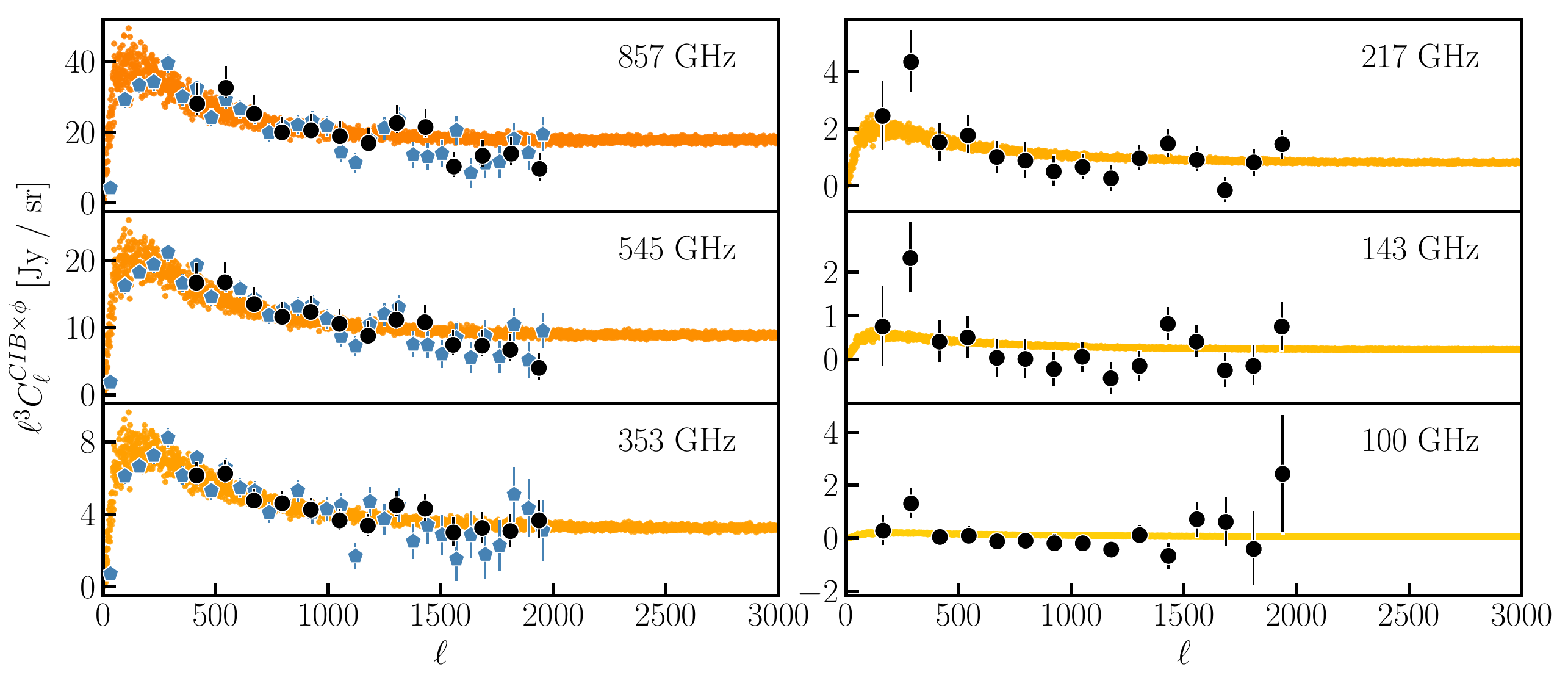}
\vspace{-0.7cm}
\end{center}
\caption{Cross correlation of the CIB and the CMB lensing potential $\phi$, including the Websky results (orange), the Planck 2013 measurements (black circles) \cite{2014A&A...571A..30P}, and the recent large-scale CIB measurements from dust-cleaned Planck maps of \cite{2019arXiv190500426L} cross correlated with the Planck lensing convergence (gray pentagons).}
\label{fig:cib-phi-power}
\end{figure}
%END FIGURE ----------------

The contribution of matter fluctuations of a given size to the CMB gravitational lensing potential $\phi$, related to the lensing convergence $\kappa$ through
\beq
\kappa = -\frac{1}{2}\nabla^2 \phi,
\eeq
peaks about half way ($z \approx 2$) between us and the last scattering surface. Similarly, the CIB redshift distribution peaks approximately around $z \simeq 2$ (with a dependence on observation frequency, being biased towards higher redshift at lower frequencies), and the dusty star-forming galaxies comprising it tend to reside in halos of mass $10^{10} < M/M_\odot < 10^{13}$. Therefore, a significant correlation is expected between the two maps \cite{2003ApJ...590..664S,2013ApJ...771L..16H}. Observationally, detecting this correlation can be used to constrain models of the CIB, while here we use it to show that the Websky simulations properly encapsulate the correlation between maps necessary to provide accurate mocks for CMB analyses. This is not guaranteed from the power spectrum results of the Websky CIB and lensing convergence independently. 

The Figure~\ref{fig:cib-phi-power} shows the cross correlation of the Websky CIB and gravitational potential maps (orange), the Planck 2013 results (black) of \cite{2014A&A...571A..30P} which used the gravitational potential reconstructed in \cite{2014A&A...571A..18P}, and the recent large-scale CIB measurements from dust-cleaned Planck maps of \cite{2019arXiv190500426L} cross correlated with the Planck lensing convergence (gray pentagons). Due to the approximately $\ell^{-1}$ and $\ell^{-2}$ shape of the CIB and gravitational potential power spectra, respectively, the cross correlation $C_\ell^{CIB \times \phi}$ is typically plotted with a leading factor of $\ell^3$, as is shown here. We find good agreement at all frequency channels for the range of $\ell$ measured by Planck, which did not consider modes with $\ell<100$ or $\ell>2000$ in their analysis due to possible mean-field systematic effects and extragalactic foreground contamination, respectively. Correlated noise between channels can enter in a number of ways from, e.g., foregrounds in multiple channels, meaning that the measurements and errorbars at each frequency channel are not completely independent, as evidenced by first two Planck data points at low frequencies. We also show the recent dust-cleaned Planck maps of Lenz et al. \cite{2019arXiv190500426L}. Again we show the baseline results, which used a low $\rm{H_I}$ column density mask corresponding to a threshold of $N_{\rm HI} = 2.5\times 10^{20}\ {\rm cm}^{-2}$. We find good agreement for both the amplitude and shape of the Websky and Lens et al. cross correlation, extending far below the original minimum Planck cutoff at these frequencies of $\ell \simeq 400$. The turn-over at low $\ell$ also matches very well, which was not possible to determine from the Planck analysis alone.

\subsubsection{CIB-tSZ Cross-correlation}
\begin{figure}[t]
\begin{center}
\includegraphics[width=0.98\textwidth, trim = 0 0 0 0, clip = True]{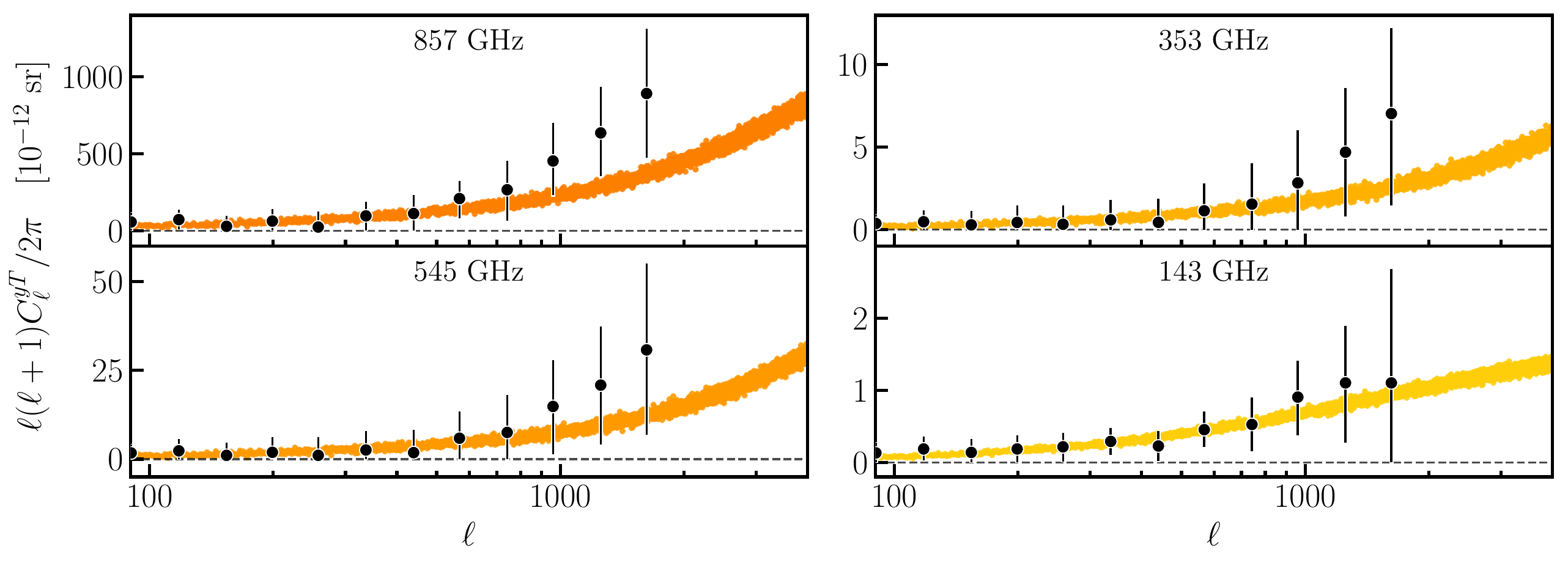}
\vspace{-0.7cm}
\end{center}
\caption{Cross correlation of the tSZ and `sky frequency maps' $T$ from 143 to 857 GHz in units of Compton-y, where a frequency map is constructed by adding the CIB and Compton-y. We show the Websky results (orange) and Planck 2015 CIB-cleaned results  \cite{2016A&A...594A..23P} (black circles). The Planck uncertainties are dominated by foreground residuals, and thus they are highly correlated from one channel to another.}
\label{fig:cib-tsz-power}
\end{figure}

The tSZ effect, unlike the CIB and gravitational lensing, is mainly produced by local high mass clusters. Therefore, the tSZ and CIB should have a much smaller overlap in redshift and halo mass than the CIB-lensing presented in the previous section. This small overlap makes the correlation hard to detect in practice, but determining the strength of the correlation remains important to determine the CIB contamination of tSZ estimates, to estimate the power resulting from the kSZ, and to constrain halo model approaches. In Figure~\ref{fig:cib-tsz-power} we show the CIB-tSZ cross correlation of the Websky maps (coloured) and the Planck 2015 results of \cite{2016A&A...594A..23P} constructed by performing a cross spectrum with the Planck frequency maps and a reconstructed MILCA y-map derived from component separation, and subtracting out the CIB leakage estimated from simulations (see \cite{2016A&A...594A..23P} for more details). The Planck data at 217 GHz was not available. The intensity units for the cross correlation, $C_\ell^{yT_\nu}$,  are ``Compton $y$-units'' \cite{2014A&A...571A..29P}, defined for each frequency bandpass $i$ such that 
\beq
y_i = I_{\nu_i} / |d\Delta{I_\nu}/dy|_i,
\eeq
with conversion factors taken from Table 6 of \cite{2014A&A...571A...9P}. See \cite{2016A&A...594A..23P} for more details on how the cross-correlation was calculated. We find broad agreement at all frequencies and multipoles for which there is significant signal-to-noise in the Planck measurements. At low frequencies the cross correlation is nearly completely determined by the Compton-y component due to the relatively low amplitude of the CIB, and we see the same turn-over at high $\ell$ as the tSZ power spectrum presented in Figure~\ref{fig:tsz-power}. By 353 GHz the CIB component begins to contribute roughly equal power as the tSZ, and by 545 and 857 GHz it dominates. See \cite{2016A&A...594A..23P} for a discussion of limitations due to systematic effects and the statistical significance of the cross-correlation as measured by the Planck Collaboration.

%BEGIN FIGURE -------------
%END FIGURE ----------------

% ===========================
% Discussion
\section{Discussion}
\label{sec:discussion}
In this paper we have described a new pipeline for generating simulated maps of the microwave sky and catalogs of galaxy groups and clusters. Because it explicitly models the fate of individual connected `patches' in Lagrangian space as a function of local measurements involving the strain of the linear displacement field, it is particularly well suited to connect observables efficiently with the initial conditions, without resorting to more costly N-body simulations. The pipeline consists of two main steps: first generate an accurate realization of large-scale structure on the past light cone and then use this realization to determine the observed spectral intensity and lensing convergence for each pixel in the simulated map. While the pipeline is applicable to a broad range of redshifts and scales, we have focused on a particular large-volume light-cone realization constructed by placing the observer at the eight corners of a periodic volume ($\sim 5.25\ {\rm Gpc}/h$ on a side with $N=6,144^3$ resolution elements), resulting in a single realization with the observer at the center of a cube $\sim 10.5\ {\rm Gpc}/h$ on a side with $N=12,288^3$ resolution elements, from which the inscribed sphere was used to generate full-sky maps and halo catalogs out to a redshift $z = 4.6$. This simulation resolved halos above a mass of approximately 10$^{12}$M$_\odot$, sufficient to include the complete tSZ signal without considering a contribution from the field component, although some signal from the low-mass end of the CIB, which is generally considered to have a minimum contributing halo mass of 10$^{10}$--10$^{11}$M$_\odot$, is missed. The free normalization parameter in the Planck CIB model used allowed for the Websky CIB maps to nevertheless reproduce the proper clustering statistics, but the omission of these low mass halos will change the average CIB bias, and is a likely reason that the Websky CIB power spectrum does not match the Planck results as well as a model that includes halos with mass less than $\sim 10^{12}\ M_\odot/h$. This signal could be included through the additional use of a field prescription, but was not included in this work. We also note that the inclusion of radio sources for the same halo catalog presented here is the subject of  work in preparation.

Spherically symmetric halo profiles were used in all the maps presented here. Since in reality individual groups and clusters are not spherical, this is a limitation that should be kept in mind when using these maps. While the halo catalogs upon which the maps are based do not contain direct information about the final state (Eulerian) orientation, the measurements performed in the mass-Peak Patch method already contain a wealth of information of Lagrangian measurements around each halo on a number of scales. This information can be used to  estimate of the final shape, orientation, and internal properties, and is a natural extension of our approach. Although we assumed the interiors of halos  to be spherical, the halo clustering and associated matter and thermal energy distribution anisotropies in the two-halo regime are accurately represented in these simulations. 

Additional tests of the mass-Peak Patch simulation method we used to construct the halo catalogues have been shown to accurately reproduce higher order halo statistics such as the halo bispectrum \cite{2019MNRAS.482.4883C}. Higher order spectra of the Websky CIB maps have also been presented in \cite{2019arXiv190502084F}, while correlations between the halo distribution and both CIB and Compton-$y$ maps were shown in \citep{2019PhRvD.100f3519P}. The map power-spectrum results and cross-spectrum results were compared to available data from Planck \cite{2014A&A...571A..30P,  2016A&A...594A..22P, 2016A&A...594A..23P}, Planck external re-analyses \cite{2018MNRAS.477.4957B, 2019arXiv190500426L}, ACT \cite{2013JCAP...10..060S}, SPT \cite{2015ApJ...799..177G}, the Herschel Multi-tiered Extragalactic Survey (HerMES) \cite{2013ApJ...772...77V}, and a number of hydrodynamical simulation results  \cite{2010ApJ...725...91B, 2012ApJ...756...15S, 2017MNRAS.467..985R, 2011ApJ...727...94T, 2016MNRAS.463.1797D, 2018ApJ...853..121P} with broad consistency found for the following validation statistics: power spectra of tSZ, kSZ, CIB, and the lensing convergence; the CIB -- CMB lensing cross-correlation;  and the CIB -- tSZ cross-correlation. The halo catalogue and maps presented here are publicly available online\footnote{ \url{mocks.cita.utoronto.ca}} and are included as part of the \texttt{so\_pysm\_models} framework\footnote{\url{https://github.com/simonsobs/so\_pysm\_models}} being developed by the Simons Observatory Collaboration.

%============================
% Acknowledgments
\acknowledgments
The majority of this work was performed while GS was a PhD candidate at the Canadian Institute of Theoretical Astrophysics. We thank Philippe Berger for many discussions on the Peak Patch method, and Amir Hajian for contributing python interfaces for early versions of the code used to make the CIB maps in this work.

Research in Canada is supported by NSERC and CIFAR. These calculations were performed on the Niagara supercomputer at the SciNet HPC Consortium \cite{2019arXiv190713600P}. SciNet is funded by: the Canada Foundation for Innovation under the auspices of Compute Canada; the Government of Ontario; Ontario Research Fund - Research Excellence; and the University of Toronto. NB acknowledges the support from the Lyman Spitzer Fellowship. AvE acknowledges the support of the Beatrice and Vincent Tremaine Fellowship at CITA.

\appendix
\section{Unit Conversions}
\label{sec:unitappendix}
Here we give the necessary factors to convert the intensity variations in CIB and thermal SZ maps to a deviation from the mean temperature of the CMB, $\rm{\Delta T}$, in thermodynamic temperature units such as $\mu K_{\rm cmb}$. The CIB intensity is typically expressed in units such as MJy sr$^{-1}$, where $\rm{Jy = 10^{-26}\ W\ m^{-2}\ Hz}$, while the thermal SZ effect is described by the dimensionless Compton-$y$ parameter. Thermodynamic temperature is used to express intensity variations on the same scale as those coming from small-amplitude blackbody temperature fluctuations, for which the derivative of the blackbody $B_\nu$ with respect to temperature evaluated at the CMB temperature, which we take to be $\rm{T_{\rm cmb}=2.7255\ K}$, is the frequency-dependent conversion factor:

\begin{align} \label{eq:Mjysr2muK}
\Delta T &= \left( \frac{dB_\nu}{dT} \right)^{-1} \Delta I_\nu 
   = \left[\frac{2 h}{c^2} \frac{\nu^3}{T_{\rm cmb}} \frac{xe^x}{(e^x-1)^2} \right]^{-1} \Delta I_\nu \\
    &= 1.05 \times 10^{3}\ \mu{\rm K_{\rm cmb}}\ (e^x-1)^2 e^{-x}\left( \frac{\nu}{100\ \rm{GHz}} \right )^{-4}   \left( \frac{\Delta I_\nu}{\rm{M Jy\ sr}^{-1}} \right) ,
\end{align}
where $x = h \nu/(k_b T_{\rm cmb}) = \nu/56.8\ \rm{GHz}$. This results in the conversion factors listed in the central column of Table~\ref{tab:conversions} needed to convert the Websky CIB maps from $\rm{M Jy\ sr^{-1}}$ to  $\rm{\Delta T_{CIB} [\mu K_{\rm CMB}]}$. 

The Compton-y parameter can be described as a change to the primordial blackbody through:
\begin{align}
    \Delta T &= y T_{\rm cmb} \left[ x \frac{e^x + 1}{e^x- 1} -4 \right],
\end{align}
where x is the same as above. Using $\rm{T_{\rm cmb}=2.7255\ K}$, this results in the conversion factors listed in the right column of Table~\ref{tab:conversions}, needed to convert the Websky Compton-y map to $\rm{\Delta T_y [\mu K_{\rm CMB}]}$.

\begin{table}
\begin{center}
\begin{tabular}{ c|c|c } 

Frequency [GHz] & [MJy sr$^{-1}$] $\rightarrow$ [$\mu$ K] & Compton-y $\rightarrow$ $[\mu {\rm K}]$ \\
\hline
27 & 4.5495 $\rm{\times 10^4}$ & -5.3487 $\rm{\times 10^6}$ \\ 
39 & 2.2253  $\rm{\times 10^4}$ &  -5.2384 $\rm{\times 10^6}$ \\ 
93 & 4.6831  $\rm{\times 10^3}$ & -4.2840 $\rm{\times 10^6}$ \\ 
100 & 4.1877 $\rm{\times 10^3}$ & -4.1103 $\rm{\times 10^6}$ \\ 
143 & 2.6320 $\rm{\times 10^3}$ & -2.8355 $\rm{\times 10^6}$ \\ 
145 & 2.5947 $\rm{\times 10^3}$ & -2.7685 $\rm{\times 10^6}$ \\ 
217 & 2.0676 $\rm{\times 10^3}$ & -2.1188 $\rm{\times 10^4}$ \\ 
225 & 2.0716 $\rm{\times 10^3}$ & 3.1517 $\rm{\times 10^5}$ \\ 
280 & 2.3302 $\rm{\times 10^3}$ & 2.7314 $\rm{\times 10^6}$ \\ 
353 & 3.3710 $\rm{\times 10^3}$ & 6.1071 $\rm{\times 10^6}$ \\ 
545 & 1.7508 $\rm{\times 10^4}$ & 1.5257 $\rm{\times 10^7}$ \\ 
857 & 6.9653 $\rm{\times 10^5}$ & 3.0228 $\rm{\times 10^7}$ 
\end{tabular}
\caption{Frequency dependent conversion factors for CIB maps in MJy sr$^{-1}$, and Compton-y maps, to a deviation $\rm{\Delta T\ [\mu K]}$ from the primordial CMB blackbody of $\rm{T_{\rm cmb}=2.7255\ K}$. These conversion factors assume an infinitely narrow passband centered at the given frequency. Multiply a CIB or Compton-y map by the given value to convert it.}
\end{center}
\label{tab:conversions}

\end{table}

%============================
% Bibliography
\bibliography{main}

\end{document}